\documentclass[sigconf]{acmart}
\AtBeginDocument{%
  \providecommand\BibTeX{{%
    \normalfont B\kern-0.5em{\scshape i\kern-0.25em b}\kern-0.8em\TeX}}}

\setcopyright{acmcopyright}
\copyrightyear{2021}
\acmYear{2021}
\acmDOI{10.1145/1122445.1122456}

\acmConference[SC'21]{SC'21: The International Conference for High Performance Computing, Networking, Storage, and Analysis}{Nov 14--19, 2021}{St.~Louis, MO}
\acmBooktitle{SC'21: The International Conference for High Performance Computing, Networking, Storage, and Analysis, Nov 14--19, 2021, St.~Louis, MO}
\acmPrice{15.00}
\acmISBN{978-1-4503-XXXX-X/18/06}

\usepackage{amsmath}
\usepackage{graphicx}
\usepackage{bm}
\usepackage{physics}
\usepackage{siunitx}
\usepackage{mathtools} % for mathclap
\usepackage[capitalise]{cleveref}
\crefname{equation}{Eq.}{Eqs.}
\crefname{figure}{Fig.}{Figs.}
\crefname{tabular}{Tab.}{Tabs.}
\crefname{section}{Sec.}{Secs.}
\crefname{chapter}{Chap.}{Chaps.}
\Crefname{equation}{Equation}{Equations}
\Crefname{figure}{Figure}{Figures}
\Crefname{tabular}{Table}{Tables}
\Crefname{section}{Section}{Sections}
\Crefname{chapter}{Chapter}{Chapters}
\usepackage{todonotes}
\usepackage{csquotes}
\usepackage{caption}
\usepackage{subcaption}

\newcommand{\supermuc}{\emph{SuperMUC-NG}}
\newcommand{\mahti}{\emph{Mahti}}
\newcommand{\shaheen}{\emph{Shaheen-II}}

\newcommand{\gigaflops}{GFLOPS}
\newcommand{\teraflops}{TFLOPS}
\newcommand{\petaflops}{PFLOPS}
\newcommand{\samoaflash}{sam(oa)\textsuperscript{2}-flash}

\begin{document}

%%
%% The "title" command has an optional parameter,
%% allowing the author to define a "short title" to be used in page headers.
%\title{3D Acoustic-Elastic Coupling for Tsunami-Genesis}
\title{3D Acoustic-Elastic Coupling with Gravity: The Dynamics of the 2018 Palu, Sulawesi Earthquake and Tsunami}
%3D Acoustic-Elastic Coupling:
%The Dynamics of the 2018 Palu, Sulawesi Earthquake-Tsunami Interaction} 
%%
%% The "author" command and its associated commands are used to define
%% the authors and their affiliations.
%% Of note is the shared affiliation of the first two authors, and the
%% "authornote" and "authornotemark" commands
%% used to denote shared contribution to the research.

% https://www.acm.org/publications/taps/latex-best-practices
% Using one author block per author is recommended..
\author{Lukas Krenz}
\orcid{0000-0001-6378-0778}
\email{lukas.krenz@in.tum.de}
\affiliation{%
  \institution{Technical University of Munich}
  \streetaddress{Boltzmannstr.\ 3}
  \city{Garching}
  %\state{Bavaria}
  \country{Germany}
  \postcode{85748}
}

\author{Carsten Uphoff}
\email{uphoff@geophysik.uni-muenchen.de}
\author{Thomas Ulrich}
\email{ulrich@geophysik.uni-muenchen.de}
\author{Alice-Agnes Gabriel}
\email{gabriel@geophysik.uni-muenchen.de}
\orcid{0000-0003-0112-8412}
\affiliation{%
  \institution{Ludwig-Maximilians-Universität München}
  \streetaddress{Theresienstr.\ 41}
  \city{Munich}
  %\state   {Bavaria}
  \country{Germany}
  \postcode{80333}
}

\author{Lauren S. Abrahams}
\email{labraha2@stanford.edu}
\orcid{0000-0002-0464-0457}
\author{Eric M. Dunham}
\orcid{0000-0003-0804-7746}
\email{edunham@stanford.edu}
\affiliation{%
  \institution{Stanford University}
  \streetaddress{397 Panama Mall}
  \city{Stanford, CA}
  \postcode{94305}
  \country{USA}
}

\author{Michael Bader}
\email{bader@in.tum.de}
\affiliation{%
  \institution{Technical University of Munich}
  \streetaddress{Boltzmannstr.\ 3}
  \city{Garching}
  %\state{Bavaria}
  \country{Germany}
  \postcode{85748}
}

%%
%% By default, the full list of authors will be used in the page
%% headers. Often, this list is too long, and will overlap
%% other information printed in the page headers. This command allows
%% the author to define a more concise list
%% of authors' names for this purpose.
\renewcommand{\shortauthors}{Krenz et al.}

%%
%% The abstract is a short summary of the work to be presented in the
%% article.
\begin{abstract}
We present a highly scalable 3D fully-coupled Earth \& ocean model of earthquake rupture and tsunami generation and perform the first fully coupled simulation of an actual earthquake-tsunami event and a 3D benchmark problem of tsunami generation by a mega\-thrust dynamic  earthquake rupture. Multi-petascale simulations, with excellent performance demonstrated on three different platforms, allow high-resolution forward modeling. % fusing tsunami, seismological, geodetic, tectonic and field observations. 
Our largest mesh has $\approx$261 billion degrees of freedom, resolving at least  \SI{15}{\hertz} of the acoustic wave field.
We self-consistently model seismic, acoustic and surface gravity wave propagation in elastic (Earth) and acoustic (ocean) materials sourced by physics-based non-linear earthquake dynamic rupture, thereby gaining insight into the tsunami generation process without relying on approximations that have previously been applied to permit solution of this challenging problem.
Complicated geometries, including high-resolution bathymetry, coastlines and segmented earthquake faults are discretized by adaptive unstructured tetrahedral meshes. This inevitably leads to large differences in element sizes and wave speeds which can be mitigated by ADER local time-stepping and a Discontinuous Galerkin discretization yielding high-order accuracy in time and space.
%A Discontinuous Galerkin discretization with ADER local time-stepping (ADER-DG) yields petascale computational efficiency and high-order accuracy in time and space.
%We compare our new 3D fully coupled approach to the existing, approximate, one-way earthquake-to-tsunami linking procedure in a 3D benchmark problem of tsunami generation by a megathrust dynamic  earthquake rupture.
%We present a large-scale fully-coupled model of the 2018 Sulawesi event that links the dynamics of supershear earthquake faulting to elastic and acoustic waves in Earth and ocean to tsunami gravity wave propagation in the narrow Palu Bay.
%And we demonstrate scalability and performance of the MPI+OpenMP parallelisation on three petascale supercomputers.
\end{abstract}

%%
%% The code below is generated by the tool at http://dl.acm.org/ccs.cfm.
%% Please copy and paste the code instead of the example below.
%%

%%
%% Keywords. The author(s) should pick words that accurately describe
%% the work being presented. Separate the keywords with commas.

%%
%% This command processes the author and affiliation and title
%% information and builds the first part of the formatted document.
\maketitle%% A "teaser" image appears between the author and affiliation
%% information and the body of the document, and typically spans the
%% page.
\begin{figure*}[h]
  \includegraphics[width=\textwidth]{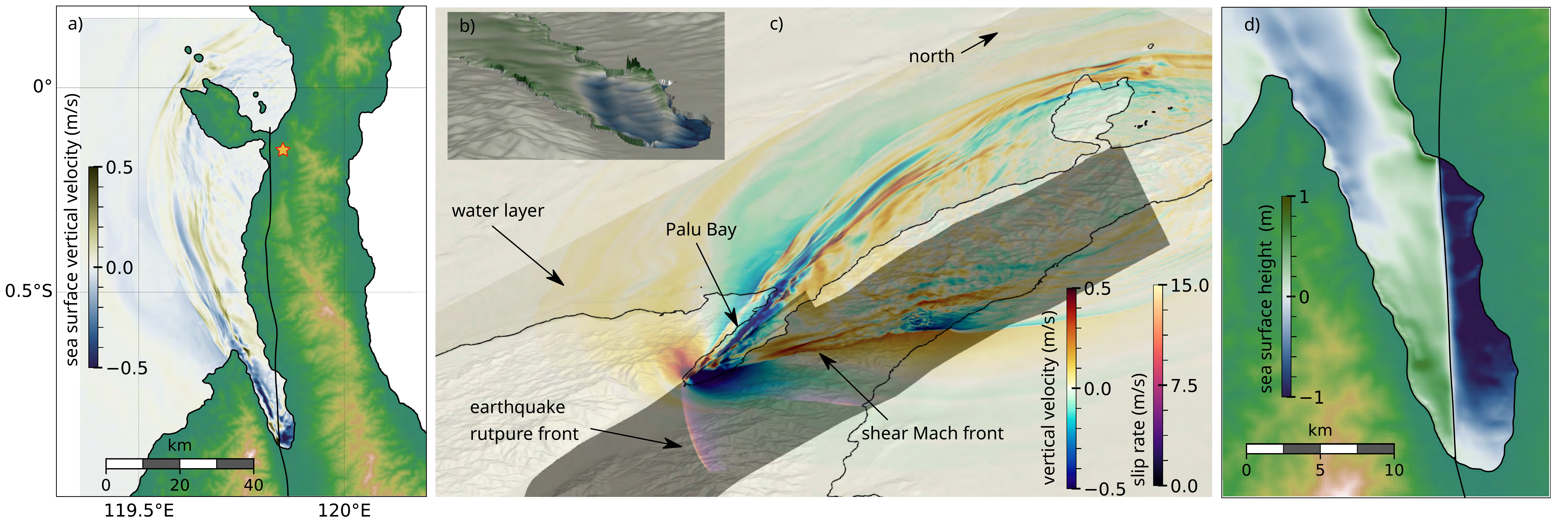}
  \caption{3D fully coupled acoustic-elastic model with gravity of the 2018 Sulawesi supershear earthquake and local tsunami in Palu Bay.
  (a) Map view of the vertical sea surface velocity at 15\,s simulation time. 
  The star indicates the earthquake epicenter. Black lines mark the traces of a system of complex faults. 
  (b) Sea surface height (m, same color map as in (d)) at 60\,s (warped vertical displacements in the bay, amplified $\times$2000).  
  (c) 3D view of earthquake and tsunami model in Palu Bay. Particle velocity (slip rate) across the faults, vertical sea-surface and Earth ground velocity at 15\,s. Dynamic earthquake rupture, propagating faster than the shear-wave velocity of the surrounding rocks, generates a clearly visible Mach cone altering the acoustic response of the ocean. 
  (d) Sea surface height (vertical displacements in Palu Bay) at 15\,s.}
%  \Description{Should be }
  \label{fig:teaser}
\end{figure*}

\section{Introduction}
\Cref{fig:teaser} illustrates a 3D fully coupled model of the tsunami caused by the magnitude 7.5 strike-slip supershear earthquake that struck Palu Bay of the island of Sulawesi, Indonesia, on September 28, 2018.
Tsunamis occur due to abrupt vertical perturbations to the water column. 
Devastating tsunamis are therefore rarely caused by strike-slip earthquakes (such as the Sulawesi earthquake), as these produce predominantly horizontal displacements. 
What was additionally unusual about the Sulawesi earthquake was its ability to produce very fast (\enquote{supershear}) movements during the rupture itself.
This surprising earthquake-tsunami behavior has led to an extensive debate with important implications for submarine strike-slip fault systems worldwide
(\citet{maiSupershearTsunamiDisaster2019} and references therein). However, efforts to understand this behavior have been hampered by the necessity of making approximations in the fluid-solid coupling, and hence in the tsunami generation process.

Recent simulations that link physics-based earthquake dynamics and elastic (seismic) wave propagation to shallow water models of tsunami genesis and propagation~\cite{ulrich_coupled_2019} suggest that the Palu tsunami can be primarily understood from the complex interaction of earthquake rupture 
%(fast \enquote{supershear} rupture propagation) 
and the 3D bathymetry of the bay, and that, while the fast rupture speed strongly influences seismic radiation and ground shaking from the earthquake, it seemingly does not affect the tsunami response.   
However, the standard practice of linking a 3D earthquake simulation, conducted without a water layer, to a 2D tsunami shallow water model simplifies the interaction between the ocean floor and water column~\cite{madden_linked_2020}.

Typical limitations of the one-way linking procedure and use of a shallow water model include neglecting dispersion (which can be somewhat addressed using Boussinesq type equations~\cite{Shi2012}), neglecting horizontal momentum transfer~\cite{Song2008,lotto_should_2017} and assuming that the ocean is incompressible~\cite{berger2011geoclaw}.
The latter assumption eliminates ocean acoustic waves, which, as our modeling demonstrates for the first time for an actual earthquake-tsunami event, can create perturbations of the ocean surface of comparable amplitude to tsunami waves but at much higher frequencies. Furthermore, at early times in tsunami generation, acoustic waves and tsunami gravity waves are superimposed, creating a complex wavefield that can be measured with ocean bottom pressure sensors and off-shore cabled sensor networks deployed in the earthquake source region with potential for improving tsunami early warning~\cite{KozdonDunham2014,Sladen2019,Zhan2021}.

% Add here motivation for fully coupled modeling
In this paper, we present a 3D fully-coupled elastic-acoustic model of the solid Earth and ocean response, with gravity, to earthquake dynamic rupture. 
Implemented in the earthquake simulation package SeisSol (www.seissol.org), the model rigorously, and without the approximations employed in one-way linking procedures, captures the entire process from earthquake rupture to the generation and propagation of seismic waves, ocean acoustic waves, and tsunamis.
We present a large-scale fully-coupled model of the 2018 Sulawesi event that links the dynamics of supershear earthquake faulting to elastic and acoustic waves in Earth and ocean to tsunami gravity wave propagation in the narrow Palu Bay. We demonstrate scalability and performance of the MPI+OpenMP parallelization on three petascale supercomputers.
The achieved efficiency of our approach allows high-resolution, data-integrated forward modeling: the dynamic rupture earthquake model fuses seismological, geodetic, tectonic and field observations and the tsunami model matches wave gauge data and inundation mapping \citep{ulrich_coupled_2019}. 

Within the heterogeneous solid Earth, we solve the elastic wave equation coupled to non-linear frictional sliding across a complex fault network.
Simultaneously, we solve the acoustic wave equation, describing perturbations about an equilibrium hydrostatic state, in a compressible, inviscid ocean of variable depth. The effects of gravitational restoring forces, which are responsible for tsunami propagation, are efficiently incorporated through a modification of the standard free surface boundary condition at the evolving sea surface~\cite{lotto_high-order_2015}. 
The enhanced modeling capabilities and gained geophysical insight come at the cost of unforeseen implementation complexity and substantially higher computational demands.
In particular, we make the following contributions:
\begin{itemize}
\item In \cref{sec:model,sec:numerics}, we present an Arbitrary high-order DERiv\-ative Discontinuous Galerkin (ADER-DG) discretization of the first 3D fully-coupled elastic-acoustic earthquake-tsunami model.
We explain how the moving sea surface can be tracked by a linearized free surface boundary condition and discretize it efficiently.
\item The fully-coupled model is implemented within the open-source software SeisSol.
In \cref{sec:hpc}, we present node-level performance (esp.\ NUMA effects) and enhancements to SeisSol's hybrid parallelization scheme, and we discuss load-balancing challenges of the fully-coupled model.
    \item In \cref{sec:results-ascete}, we evaluate the fully-coupled model in a 3D benchmark setup of tsunami generation by a dynamic earthquake rupture, and compare against a shallow water one-way linking simulation approach.
    \item We simulate a large-scale 3D fully-coupled elastic-acoustic scenario of the 2018 Palu earthquake and tsunami (\cref{sec:results-palu})
    and show that we capture salient features of tsunami genesis.
    Our largest setup has 518 million elements (261 billion degrees of freedom).
    We achieve sustained petascale performance for this setup on two supercomputers, \shaheen{} and \supermuc{}.
    \item In \cref{sec:results-scaling}, we demonstrate that our method achieves a parallel efficiency of $>$70\% on the \mahti{} and \supermuc{} supercomputers (based on AMD and Intel architectures, from 50 to 700 and from 50 to 1600 nodes, respectively), examining the strong-scaling limit and focusing on NUMA-aware allocation of processes to nodes.
\end{itemize}
\section{Related Work}\label{sec:related}
Tsunami modeling typically uses one-way linking of the earthquake-induced seafloor uplift to the ocean response. This is mostly justified when the earthquake happens so quickly that the tsunami propagation distance during the rupture duration is much smaller than the source dimension. In this approximation, the final, static seafloor uplift is utilized as an initial condition for the tsunami. The seafloor uplift is furthermore commonly simplified by using analytical solutions for the displacement caused by fault slip within a homogeneous elastic half-space (\citet{okada_surface_1985}). The long-wavelength components of the seafloor uplift are then assumed to instantaneously uplift the water column.
In reality, the seafloor moves dynamically in response to sources such as earthquakes and landslides, generating seismic waves, ocean acoustic waves and tsunami gravity waves. While these waves eventually separate at a sufficiently large distance from the source, they are superimposed in the source region. Offshore instrument deployments in source regions of potential events have motivated the development of methods that capture the full wavefield.
\citet{Maeda2013} developed the first such coupled model in their study of the 2011 Tohoku, Japan, event, utilizing 3D kinematic earthquake sources.
%\citet{uphoff_extreme_2017} and %\citet{ulrich_stress_2020} simulate the dynamic rupture process of the 2004 Sumatra-Andaman earthquake and link the simulated ocean floor displacement to a 2D tsunami propagation model.  
\citet{Saito2019} present an alternative method in which the full wavefield is constructed as the superposition of a solution conducted without gravity, but including a compressible ocean, and a second solution to the tsunami problem using forcing from the zero-gravity solution.

Another approach, which we take in this work, is to directly solve the linearized equations of motions governing smaller perturbations of the ocean about a rest state in hydrostatic equilibrium, with rigorous acoustic-elastic coupling to the solid Earth across the deforming seafloor interface. Thus far, this approach, which involves a modification of the free surface condition, has only been realized in 2D ~\cite{lotto_high-order_2015,lotto_should_2017,lotto2018fully}.
These generic 2D studies reveal the fundamentally complex, dynamic nature of the tsunami generation process, which superimposes the initial tsunami waveforms with ocean acoustic waves, oceanic Rayleigh waves, and other seismic waves in simple model setups.
%The fully-coupled method allows for realistic geometries and rupture processes, providing a deeper understanding of the relationships between seismic and acoustic amplitudes and periods and tsunami heights, and also serves as a reference solution to which simpler, approximate methods can be compared.
Of course, this method must be extended to 3D in order to assimilate and interpret observational data, to better understand the earthquake rupture and tsunami generation process and to improve tsunami early warning capabilities. 
The fully-coupled method allows for realistic geometries and rupture processes, providing a deeper understanding of the relation between seismic and acoustic amplitudes and periods and tsunami heights, and also serves as a reference solution to which simpler, approximate methods can be compared.

3D high-resolution simulation of earthquakes (including tsunamigenic events) and seismic shaking allow geophysical insight and mitigation of natural hazards, even when complicated by sparse data coverage \cite{Milner2021}. 
Integration with geophysical and geological constraints, such as complex structure and topography, requires massive computational performance, demonstrated in several petascale simulation efforts during the last decade. 
For example, \citet{Roten:SC16} present non-linear ground motion models with off-fault plasticity on the Blue Waters and Titan supercomputers;  
\citet{Fu:SC17} and \citet{chen:SC18} model the Wenchuan Earthquake, incorporating high-resolution surface topography on Sunway TaihuLight. 
Recent work using the SeisSol software integrates dynamic rupture models of tsunamigenic earthquakes with seismic, geodetic and tsunami data \cite{uphoff_extreme_2017,ulrich_coupled_2019,ulrich20}. 
\citet{ichimura:SC14,ichimura:SC18} present coupled finite-element simulations of earthquake-induced ground shaking and their impact on urban structures.
\citet{Rodgers2018, Rodgers2020} perform broadband anelastic ground motion scenario simulations in the San Francisco Bay Area.

\section{Model Description}\label{sec:model}
In the following, we lay out the 3D elastic-acoustic model, including interface and boundary conditions, used for our simulations. 
We use Einstein summation convention, i.e., summation over indices appearing twice is implied.

\textit{Earth and rupture model:}
In the solid Earth, we solve the seismic wave equation in velocity-stress formulation.
With the unknowns\\ $\bm{q} = \left( \sigma_{xx}, \sigma_{yy}, \sigma_{zz}, \sigma_{xy}, \sigma_{yz}, \sigma_{xz}, v_x, v_y ,v_z\right)$, i.e., six stress and three velocity components,
and using indices $i,j \in \{ x,y,z \}$, we can state our set of equations as
\begin{align}\label{eq:elasticity-pde}
\begin{split}
\pdv{\sigma_{ij}}{t} - \lambda \delta_{ij} 
\pdv{v_k}{x_k}
- \mu \left(\pdv{v_i}{x_j} + \pdv{v_j}{x_i} \right) &= 0,\\
\rho \pdv{v_i}{t} - \pdv{\sigma_{ij}}{x_j} &= 0.
\end{split}
\end{align}
Here, $\mu, \lambda$ and $\rho$ are the spatially-varying material properties and $\delta_{ij}$ is the Kronecker delta.

An earthquake in the domain $\Omega_E$ is modeled as frictional shear failure on a pre-defined infinitesimally thin but potentially geometrically complex fault $\Gamma_F \subset \partial\Omega_E$.
The following boundary conditions have to be satisfied on $\Gamma_F$:
\begin{align}\label{eq:bc-dynrup}
\begin{split}
    |\tau| &\leq \tau_S, \\
    \tau_S V_i - \tau_i |V| &= 0, \\
    \tau_S &= s f(|V|,\psi),  \\
    \dv{\psi}{t} &= g(|V|,\psi),
\end{split}
\end{align}
where $s=n_i\sigma_{ij}n_j$,
$\tau_i=(\delta_{ij} - n_in_j)\sigma_{jk}n_k$,
$V_i = (\delta_{ij}-n_in_j)(v_j^+-v_j^-)$,
$n$ is the unit normal pointing from the ``$-$''-side to the
``$+$''-side, and $|.|$ is the vector length.
The function $f$ returns the coefficient of friction that depends
on the slip-rate $|V|$ and the state variable $\psi$. The evolution
of $\psi$ in time is controlled by the function $g$ \cite{pelties_gmd_2014}.

\textit{Ocean model}:
Following~\citet{lotto_high-order_2015}, the ocean dynamics
are modeled as a small perturbation about the hydrostatic
state, that is, pressure is given by
$p(x,y,z,t)=p_0(z) + p'(x,y,z,t)$.
We adopt an Eulerian description with the sea level at $z=0$.
The positive $z$-axis points up.
The equations of momentum and mass balance are linearized about the hydrostatic state such that
\begin{align}\label{eq:acoustics-pde}
\begin{split}
\pdv{p'}{t} + K \pdv{v_k}{x_k} &= \rho gv_z,\\
\rho \pdv{v_i}{t} + \pdv{p'}{x_i} &= -\delta_{iz}\frac{\rho g}{K}p,
\end{split}
\end{align}
where $K$ is the bulk modulus.
Note that in \cref{eq:acoustics-pde} we use the sign convention that pressure is positive in compression; in contrast, stress is positive in tension in \cref{eq:elasticity-pde,eq:bc-dynrup}.
We identify \cref{eq:acoustics-pde} as the equations of linear acoustics, if the right-hand side is set to zero.
The terms on the right-hand are neglected because they are several orders of magnitude smaller than the terms that are retained~\cite{lotto_high-order_2015}.

The pressure on the moving sea surface must be equal to the atmospheric pressure $p_a$, i.e.,
\begin{equation}\label{eq:boundary-absolute-pressure}
    p(x, y, \eta(x,y,t),t) = p_a,
\end{equation}
where $\eta$ is the displacement of the sea surface.
By linearizing and inserting the pressure gradient in the hydrostatic state, we can simplify \cref{eq:boundary-absolute-pressure} as~\cite{lotto_high-order_2015}:
\begin{equation}
    p(x, y, 0, t) \approx
        p_a + \rho g\eta(x,y,t),
\end{equation}
where $g=\SI{9.81}{\metre\per\second^2}$.
Thus, we impose the following boundary conditions at $z=0$:
\begin{align}\label{eq:gravitational-bc}
    p'(x,y,z=0,t) &= \rho g \eta ,\\
    \pdv{\eta(x,y,z=0,t)}{t} &= u_z(x,y,z=0,t).
    \label{eq:eta-ode}
\end{align}
In this way, our model incorporates gravitational effects via a modification of the surface boundary condition, which is applied on the equilibrium ocean surface, thereby avoiding the need for a time-dependent mesh that tracks the moving surface.

In summary, our 3D fully-coupled model includes earthquake rupture dynamics, wave propagation in both elastic and acoustic media, as well as tsunami propagation. The model naturally accounts for effects of compressibility, as well as non-hydrostatic ocean response that leads to dispersion and related effects during tsunami propagation and generation.

\section{Discretization \& Implementation}\label{sec:numerics}
We implement the model from \cref{sec:model} in the high-order earthquake simulation software SeisSol (e.g., \cite{heinecke_petascale_2014,uphoff_extreme_2017}). 
We first summarize the ADER-DG method used in SeisSol and explain how we extend it to include the acoustic-elastic coupling and gravitational free surface boundary condition.
Then, we explain details of SeisSol's local-time-stepping approach and optimized parallel implementation, and how we utilize these for our coupled model.
The resulting implementation is the first to allow for 3D fully-coupled elastic-acoustic simulations that capture the full dynamics of the earthquake process, seismic wave propagation, the displacement of seafloor and surface and the onset of tsunami propagation.

\subsection{The ADER-DG method}
The acoustic wave equation \eqref{eq:acoustics-pde} becomes a special case of \cref{eq:elasticity-pde} by setting $K = \lambda$, $\mu=0$, and $\sigma_{ij} = -p \delta_{ij}$. Hereafter we use $p$ to denote the pressure perturbation $p'$.
We may thus write both \cref{eq:elasticity-pde,eq:acoustics-pde} in the general non-conservative hyperbolic form
\begin{equation}\label{eq:linear-pde}
    \pdv{\bm{q}}{t} 
    + \bm{A} \pdv{\bm{q}}{x} 
    + \bm{B} \pdv{\bm{q}}{y} 
    + \bm{C} \pdv{\bm{q}}{z} 
    = 0,
\end{equation}
where $\left( \bm{A}, \bm{B}, \bm{C} \right) \in \mathbb{R}^{9 \times 9}$ are the space-dependent Jacobian matrices.
We note that embedding the acoustic equations into the equations of elasticity entails a computation and memory overhead, however, existing data-structures in our inherited code-base are unaffected and as such the incorporation of the ocean model is less invasive.

We describe the Discontinuous Galerkin (DG) method briefly in the following:
The domain $\Omega$ is approximated using conforming tetrahedral meshes $\mathcal{T}_h = \{K\}$.
We define the usual broken finite element space
\begin{equation}
    W_h := \{\bm{p}\in [L^2(\Omega)]^9 : \bm{p}|_K \in [\mathcal{P}_N(K)]^9 \quad \forall K \in\mathcal{T}_h\},
\end{equation}
where $\mathcal{P}_N(K)$ is the space of polynomials on $K$ of maximum degree~$N$.
The space-dependent Jacobians $\bm{A}, \bm{B}, \bm{C}$
are element-wise constant.
We derive the semi-discrete form of \cref{eq:linear-pde} by multiplying with a test function $\bm{p}\in W_h$, integrating over an element $K$ and by integration by parts:
\begin{multline}\label{eq:semi-discrete-form}
\int_{K} \bm{p} \cdot \pdv{\bm{q}}{t} \dd{V}
- \int_{K}
    \left(
    \pdv{\bm{p}}{x} \bm{A} 
    + \pdv{\bm{p}}{y} \bm{B} 
    + \pdv{\bm{p}}{z} \bm{C} 
    \right)
    \bm{q}
 \dd{V} \\
+ \int_{\partial K} \bm{p}\cdot (n_x\bm{A}+n_y\bm{B}+n_z\bm{C})\bm{q^*} \dd{S} = 0,
\end{multline}
where $(n_x\bm{A}+n_y\bm{B}+n_z\bm{C})\bm{q^*}$ is the numerical flux, which we discuss later in more detail, $(n_x,n_y,n_z)$ is the outward-pointing normal vector.
The trial space is the same as the test space.
Thus, using Dubiner's basis $\phi_l$ for $\mathcal{P}_N(K)$
we can expand~$\bm{q}$ as
\begin{equation}
    \bm{q}(t,\bm{x}) = \sum_{l=1}^{B_N} \bm{Q}_l(t) \phi_l(\bm{x}),
    \qquad\textstyle
    \text{where $B_N = {N+3 \choose 3}$.}
\end{equation}

For the efficient implementation of \cref{eq:semi-discrete-form}, we map every element to a reference element.
The resulting integrals only depend on the reference element and are pre-computed, leading to a quadrature-free computation of stiffness and flux terms.
We refer the interested reader to \cite{dumbser_arbitrary_2006,uphoff_extreme_2017} for further details.

We employ ADER time-stepping, which is a predictor-corrector approach that results in an efficient one-step update~\cite{titarev2002ader,dumbser_arbitrary_2006,Kaser2006}.
In the predictor step, we expand the cell-local solution in time
\begin{align}\label{eq:taylor}
\bm{Q}_l(t) \approx \sum_{k=0}^N \frac{(t - t_0)^k}{k!} 
\pdv[k]{\bm{Q}_l}{t}\left(t_0\right)
\end{align}
around the beginning of our timestep $t_0$.
We compute the time derivatives of $\bm{Q}_l$ with the discrete Cauchy-Kowalesky procedure~\cite{puente_arbitrary_2007}, i.e., time-derivatives are replaced by spatial derivatives using the PDE recursively.
We obtain the corrector step by inserting \cref{eq:taylor} into the time-integrated \cref{eq:semi-discrete-form}.

\subsection{Elastic-acoustic coupling}
In \cref{eq:semi-discrete-form}, the numerical flux, i.e., the term $(n_x\bm{A}+n_y\bm{B}+n_z\bm{C})\bm{q^*}$, must be
chosen carefully to ensure consistency and stability of our ADER-DG scheme.
We choose the Godunov numerical flux, i.e., we solve the one-dimensional Riemann problem exactly for all basis functions on the faces~\cite{toro_riemann_2009,leveque_finite_2002}.
\begin{figure}
    \centering
        \begin{tikzpicture}
      \filldraw[draw=none,fill=black!5] (0,0) rectangle (2,2);
      \draw[->] (-2,0)  -- (2,0) node [right] {$x$};
      \draw[->] (0,0)  -- (0,2) node [above] {$t$};
      \draw (-1,0) node [below] {elastic};
      \draw (1,0) node [below] {acoustic};
      \draw[very thick] (0,0)  -- (2.0,1.1) node [above,right] {$c_p^+$};
      \draw[very thick] (0,0)  -- (-1.9,2.0) node [left] {$-c_p^-$};
      \draw[very thick] (0,0)  -- (-0.9,2.0) node [above] {$-c_s^-$};
      \draw[very thick] (1.5,0.15) node [above] {$\bm{q}^+$};
      \draw[very thick] (-1.5,0.15) node [above] {$\bm{q}^-$};
      \draw[very thick] (0.45,1.3) node [above] {$\bm{q}^{c}$};
      \draw[very thick] (-0.35,1.3) node [above] {$\bm{q}^{b}$};
      \draw[very thick] (-1.0,1.2) node [above] {$\bm{q}^{a}$};
    \end{tikzpicture}
    \caption{\label{fig:elastic-acoustic-riemann}%
    Solution structure of the Riemann problem in coupled elastic-acoustic media.
    The eigenstructure implies that we have one left-going P-wave and two left-going S-waves in the elastic medium.
    In contrast, only one right-going acoustic wave is present in the acoustic medium.
    }
\end{figure}
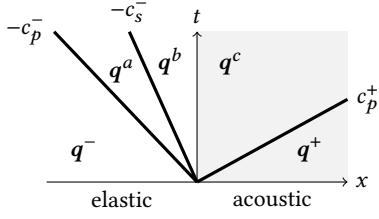

Due to the eigenstructure of \cref{eq:elasticity-pde}, we have one (compressional) P-wave and two (shear) S-waves in the elastic domain, traveling with speeds $c_p^-$ and $c_s^-$, respectively.
In the acoustic domain, we have one P-wave (acoustic wave), traveling with speed $c_p^+$.
The resulting solution structure at an elastic-acoustic interface is shown in \cref{fig:elastic-acoustic-riemann}.
The jumps $\bm{q}^a-\bm{q}^-$ and $\bm{q}^b-\bm{q}^+$ have to satisfy the Rankine-Hugoniot conditions
\begin{align}\label{eq:rankine-hugoniot}
\begin{split}
    \bm{\hat{A}}^-(\bm{q}^--\bm{q}^a) &= -c_p^-(\bm{q}^--\bm{q}^a), \\
    \bm{\hat{A}}^-(\bm{q}^a-\bm{q}^b) &= -c_s^-(\bm{q}^a-\bm{q}^b), \\
    \bm{\hat{A}}^+(\bm{q}^c-\bm{q}^+) &= c_p^+(\bm{q}^c-\bm{q}^+),
\end{split}
\end{align}
where $\bm{\hat{A}}^{\pm} = n_x\bm{A}^{\pm}+n_y\bm{B}^{\pm}+n_z\bm{C}^{\pm}$.
Note that $\bm{\hat{A}}^{-}$ denotes the rotated Jacobian of the elastic domain with Lamé parameters $\lambda^-,\mu^-$, whereas $\bm{\hat{A}}^{+}$ denotes the rotated Jacobian of the acoustic domain with bulk modulus $\lambda^+ = K^+$.

\Cref{eq:rankine-hugoniot} is an eigenproblem.
Therefore, we can state the jumps as a linear combination of eigenvectors
\begin{align}\label{eq:eigendecomposition}
    \begin{split}
        \bm{q}^--\bm{q}^b &= \alpha_1\bm{\hat{r}}_1^-+\alpha_2\bm{\hat{r}}_2^-+\alpha_3\bm{\hat{r}}_3^-, \\
        \bm{q}^c-\bm{q}^+ &= \alpha_4\bm{\hat{r}}_4^+,
    \end{split}
\end{align}
where $(-c_p^-,\bm{\hat{r}}_1^-)$, $(-c_s^-,\bm{\hat{r}}_2^-)$, $(-c_s^-,\bm{\hat{r}}_3^-)$ are eigenpairs of $\bm{\hat{A}}^-$ and $(c_p^+,\bm{\hat{r}}_4^+)$ is an eigenpair of $\bm{\hat{A}}^+$.
In practice, it is not necessary to find these eigenpairs for every instance of the normal vector $\bm{n}$, as both acoustic and elastic wave equations are rotational invariant.
The resulting similarity transform, represented by the matrix $\bm{T}$~\cite{dumbser_arbitrary_2006}, leads to the identity
\begin{align}
    \bm{\hat{A}} = n_x\bm{A}+n_y\bm{B}+n_z\bm{C}
                 = \bm{T}(\bm{n}) \bm{A}\bm{T}^{-1}(\bm{n}).
\end{align}
Thus, $\bm{\hat{r}}_i = \bm{T}\bm{r}_i$, where $\bm{r}_i$ is an eigenvector of $\bm{A}$.
Multiplying \cref{eq:eigendecomposition} with $\bm{T}^{-1}$ and letting $\bm{w} = T^{-1}\bm{q}$ gives
\begin{align}\label{eq:eigendecomposition-rotinv}
    \begin{split}
        \bm{w}^--\bm{w}^b &= \alpha_1\bm{r}_1^-+\alpha_2\bm{r}_2^-+\alpha_3\bm{r}_3^-, \\
        \bm{w}^c-\bm{w}^+ &= \alpha_4\bm{r}_4^+.
    \end{split}
\end{align}

\Cref{eq:eigendecomposition} constitutes an under-determined linear system of equations.
In order to close the system of equations, we require continuity of traction and continuity of the normal component of the velocity, as appropriate for coupling an elastic solid to an inviscid fluid~\cite{sochacki_1991}.
Here, these interface conditions are $w_i^b=w_i^c, i=1,7$ and $w_i^b = w_i^c = 0, i=4,6$.

Solving the Riemann problem gives the boundary state
\begin{align}\label{eq:boundary-state}
\begin{array}{ll}
    w_1^b = w_1^- - \alpha_1, &
    w_7^b = w_7^- - \dfrac{1}{Z_p^-}\alpha_1, \\
    w_4^b = 0, &
    w_8^b = w_8^- - \dfrac{1}{Z_s^-}w_4^-, \\
    w_6^b = 0, &
    w_9^b = w_9^- - \dfrac{1}{Z_s^-}w_6^-,
\end{array}
\end{align}
where $Z_p = c_p\rho$, $Z_s=c_s\rho$, and
\begin{align}
    \alpha_1 = \frac{Z_p^-Z_p^+}{Z_p^- + Z_p^+} \left(\frac{w_1^- - w_1^+}{Z_p^+} + w_7^- - w_7^+\right).
\end{align}
The components $w_8^b$ and $w_9^b$ are the tangential particle
velocities.
We observe that the tangential traction components $w_4^-, w_6^-$ penalize the tangential particle velocities, thus, weakly enforcing that shear tractions must be zero on
the elastic-acoustic interface.
Note that the components $w_2^b,w_3^b,w_5^b$ are irrelevant as they do not contribute to the numerical flux due to the non-zero structure of $\bm{A}$.

It is straightforward to see that the boundary state is consistent with the physical interface conditions by inserting $w_i^- = w_i^+$, $i=1,7$, and $w_i^-=w_i^+=0$, $i=4,6$ in \cref{eq:boundary-state}.
Failing to ensure consistency, e.g., if a flux is used that only takes material parameters from one side into account, as e.g., in~\cite{dumbser_arbitrary_2006,kaser_quantitative_2008}, may lead to a non-converging scheme when coupling elastics and acoustics~\cite{wilcox_high-order_2010}.

In total, the numerical flux is given by
\begin{equation}\label{eq:numerical-flux}
    \bm{\hat{A}}^-\bm{q}^* =
        \bm{T}\bm{A}^-
        \bm{w}^b(\bm{T}^{-1}\bm{q}^-,\bm{T}^{-1}\bm{q}^+).
\end{equation}
In our implementation, we exploit that $\bm{w}^b$ is bilinear and rewrite \cref{eq:numerical-flux} as
\begin{equation}
    \bm{\hat{A}}^-\bm{q}^* =
        \bm{F}^-\bm{q}^- + 
        \bm{F}^+\bm{q}^+.
\end{equation}
The $9 \times 9$ matrices $\bm{F}^-$ and $\bm{F}^+$ can be pre-computed for each face.
Thus, we can use the exact Riemann solver for little additional cost.

\subsection{Boundary conditions}\label{sec:boundary-conditions}
%As typical in discontinuous Galerkin schemes,
We impose the boundary conditions weakly, i.e., through the numerical flux.
We construct ghost values $\bm{q}^+$ such that the solution of the Riemann problem
results in the correct state at the boundary.
The solution of the \enquote{inverse Riemann problem} for boundary condition \cref{eq:gravitational-bc} is given by
\begin{align}
u_n^+ = v_n^-, \quad
p^+ = 2 \rho g \eta - p^-,
\end{align}
where $v_n$ is the velocity in normal direction.
Solving the Riemann problem with the ghost state and interior state as input, we obtain
\begin{align}\label{eq:free-surface-boundary-state}
        v_n^b = v_n^- - \frac{1}{Z} (\rho g \eta - p^-), \quad
        p^b = \rho g \eta,
\end{align}
where the superscript $b$ denotes the middle state (cf.\ \cref{fig:elastic-acoustic-riemann})
and $Z = c_p\rho$ is the impedance.
The value of $v_n^b$ can be interpreted as an extrapolation of the interior value $v_n^-$ and a penalty term that measures the difference between the pressure at the boundary~$p^-$ and the actual pressure $\rho g \eta$.

The displacement $\eta$, cf.\ \cref{eq:eta-ode},
is evolved with
\begin{align}\label{eq:free-surface-boundary-ode}
    \pdv{\eta}{t} &= v_n^b = v_n^- - \frac{1}{Z} (\rho g \eta - p^-).
\end{align}
It is critical to use the velocity $v_n^b$ here
(instead of, e.g., $v_n^-$)
as only then we have a stable scheme.

For the ADER-DG scheme, we need to integrate the boundary state in time.
We do so by integrating the coupled ODE system
\begin{align}\label{eq:coupled-ode-system}
\begin{split}
    \pdv{\eta}{t} &= v_n^- - \frac{1}{Z} (\rho g \eta - p^-), \\
    \pdv{H}{t} &= \eta,
\end{split}
\end{align}
with initial conditions for each timestep
\begin{align}
        \eta(t_n) = \eta^n, \quad
    H(t_n) = 0,
\end{align}
where $\eta^n$ is the displacement at the beginning of the timestep.
In this way, we can compute the time-integrated displacement at the same time as the displacement.
The time-integrated boundary states are obtained with
\begin{align}
        \int_{t_n}^{t_{n+1}} u_n^b\dd{t} = \eta^{n+1} - \eta^{n}, \quad
    \int_{t_n}^{t_{n+1}} p^b \dd{t} = \rho g H(t^{n+1}).
\end{align}

The time-integration of \cref{eq:coupled-ode-system} can be performed using standard ODE solvers.
We use Verner's \enquote{most efficient} Runge-Kutta scheme of order 7~\cite{VernerRungeKutta}.
In the Runge-Kutta stages, we need to evaluate the right-hand side of \cref{eq:coupled-ode-system} at several points in time.
In order to evaluate $v_n^-$ and $p^-$ at arbitrary times in the interval $[t_n, t_{n+1}]$, we use \cref{eq:taylor} to predict the evolution of $v$ and $p$ in the volume and then
extrapolate the volume data to the boundary.

Note that this implementation is more efficient than computing the analytical solution of the boundary ODE in \cref{eq:free-surface-boundary-ode} and subsequent integration.
This would require computing integrals (e.g., of the pressure) for each time quadrature point, i.e., a nested quadrature scheme.
As the evaluation of $v^-, p^-$ is the most expensive part of computing the boundary condition, this would incur significant computational costs without increasing the accuracy.

\subsection{Local time-stepping}
The stability of the explicit ADER-DG scheme is controlled by the (Courant–Friedrichs–Lewy) CFL condition \cite{courant1928,dumbser_unified_2008}
\begin{equation}
\Delta t \leq C(N) h (\vert \lambda^\text{max}\vert)^{-1},
\end{equation}
where $C(N) \leq \left(2N+1\right)^{-1}$ is a constant that depends on the polynomial order and $h$ is the maximum diameter of the insphere of the tetrahedron.
The maximum eigenvalue of the Jacobian $\lambda^\text{max} = \sqrt{\left(\lambda + 2 \mu\right)/\rho}$ is the P-wave speed (speed of sound in the acoustic layer) and thus only depends on the material parameters.
Considering geophysically motivated realistic scenarios with material heterogeneity, geometric complexity and statically adaptive meshes leads inevitably to large differences in element sizes and wave speeds which can be mitigated by local time-stepping~\cite{madec2009,rietmann2015,breuer_petascale_2016,rietmann2017,uphoff_extreme_2017}.
Our proposed acoustic-elastic coupling adds to this challenge, as the wave speeds in the Earth and the ocean differ significantly.
Furthermore, incorporating shallow water depths, as in the case of the Palu tsunami, demand locally fine spatial discretizations.
It is easy and efficient to implement local time-stepping (LTS) with ADER due to the inherent polynomial-in-time representation of our discrete solution~\cite{dumbser_arbitrary_2007}.
LTS was also shown to reduce dispersion and dissipation errors associated with unnecessarily small timesteps~\cite{dumbserArbitraryHighOrder2006,dumbser_arbitrary_2007}. 
SeisSol provides a multi-rate local time-stepping implementation, which clusters elements with similar timesteps~\cite{breuer_petascale_2016,uphoff_extreme_2017}, such that the CFL condition of no element is violated and that each element is within the cluster with the largest admissible timestep size.
Parallelization thus works over batches of elements and avoids the otherwise erratic element-wise time-stepping.
We chose rate-2 LTS, such that the first cluster contains all elements with the smallest overall timestep size, the second cluster all elements with a timestep size larger than $2 \left(\Delta t\right)_\text{min}$, and so on.
In our experiments (see \cref{sec:results}), we find that LTS has an even stronger influence on time-to-solution than reported previously for earthquake simulations, which we attribute to the discrepancy of wave speeds and mesh resolution in the acoustic and elastic layer.
See also the histogram of timestep sizes in \cref{fig:clustering-palu}. 

\section{HPC aspects}\label{sec:hpc}
Our coupled elastic-acoustic model is integrated in the software package SeisSol that is optimized for earthquake simulations at petascale~\cite{heinecke_petascale_2014,uphoff_extreme_2017}.
Here, we briefly summarize SeisSol's optimization and parallelization strategy and outline which additional contributions were necessary to realize the elastic-acoustic landmark simulations of \cref{sec:results}: (i) we demonstrate strong NUMA effects of node-level performance, especially on AMD Rome, (ii) we improve support for using multiple MPI ranks per node, via careful pinning of OpenMP threads and POSIX threads used for communication, and (iii) we implement performance-aware static load-balancing on heterogeneous supercomputers.

\subsection{Node-level performance}\label{sec:node-level-performance}

High-order ADER-DG schemes spend the majority of the computational time in element-local kernels, most of which consist of linear algebra operations of small size.
These kernels can be expressed as a sequence of tensor operations, potentially involving sparse matrices.
SeisSol implements these tensor expressions using the code generator YATeTo~\cite{uphoff_yet_2019}.
YATeTo transforms tensor expressions into sequences of small GEMMs whenever possible.

Small-GEMM subroutines are generated by hardware-specific backends, such as LIBXSMM~\cite{heinecke_libxsmm_2016} or SeisSol's  PSpaMM\footnote{\url{https://github.com/peterwauligmann/PSpaMM}}.
On the Skylake architecture, SeisSol uses both generators:
PSpaMM for dense-sparse matrix-matrix multiplications and LIBXSMM for dense-dense matrix-matrix multiplications.
On the AMD Rome architectures we use LIBXSMM and generate code for the Intel Haswell architecture which results in efficient assembler code.\footnote{PSpaMM only supports code generation for AVX512.}

Performance is evaluated by a
performance reproducer for the wave propagation part.
We expect to see NUMA effects in the corrector step due to unstructured memory access over NUMA boundaries, but not in the predictor step, which only requires element-local data.
We tested this hypothesis on a dual-socket AMD Rome 7H12 system with base frequency of \SI{2.6}{\giga \hertz}. 
%and turbo frequency of \SI{3.3}{\giga \hertz}. 
Each CPU has 64 cores and 4 NUMA nodes.
For all benchmarks the entire node is used to ensure that the boost frequency (\SI{3.3}{\giga \hertz}) is not triggered.
We thus assume a peak performance of 5325 \gigaflops{} per node.
Running only the predictor step of our scheme, we get a performance of 3360 \gigaflops{} (63\% of peak) on the entire node and a performance of 428 \gigaflops{} on a single NUMA node.
Extrapolating the result on a single NUMA node predicts a performance limit of 3424 \gigaflops{} (64\% of peak), i.e., NUMA has little effect on the predictor step.
Running both predictor and corrector step of the scheme, we get a performance of 2053 \gigaflops{} (38\% of peak) on the entire node and  376 \gigaflops{} on a single NUMA node.
Extrapolation of the latter result predicts a performance limit of 3008 \gigaflops{} (56\% of peak).
Note that we also ran this on one entire socket and get a performance of 1390 \gigaflops{} (52\% of peak).
Thus, we confirm strong NUMA effects in the corrector step on this architecture.

\subsection{Hybrid parallelization}
SeisSol is parallelized using MPI, OpenMP, and POSIX threads (pthreads).
%
%The MPI parallelization uses classical domain decomposition techniques, which requires careful load balancing.
%
On each MPI partition, OpenMP is used to parallelize loops on the level of time clusters, i.e., 
bulk-synchronous loops over elements and dynamic rupture faces are distributed with OpenMP for all elements or dynamic rupture faces in one time cluster.
In addition, we use a communication thread (implemented with pthreads) to ensure MPI progression and truly asynchronous communication~\cite{hoefler2008,breuer_petascale_2016}.
%The problem with this style of communication is that
%Communication threads are implemented with pthreads. %% see above
Pthreads are also created for asynchronous I/O~\cite{rettenberger_optimizing_2015}.

The default operation mode of SeisSol has been to use one rank per node with one communication thread and NUMA-aware memory initialization.
This mode of operation resulted in best performance on previous generation supercomputers, but the results in \cref{sec:node-level-performance} suggest that multiple ranks per node increase performance on systems with many NUMA domains.

Communication (and I/O) threads must not be pinned to worker threads, as the heavy use of \texttt{MPI\_Test} within the communication thread would interfere with the worker.
We develop a simple and portable algorithm to correctly pin communication threads when using multiple ranks per node.
Affinity of worker threads is controlled via the \texttt{OMP\_PLACES} and \texttt{OMP\_PROC\_BIND} environment variables.
We set the number of OpenMP threads to leave one physical core per MPI rank unused.
Next, we pin communication threads to free cores using \texttt{sched\_setaffinity}.
To determine free cores,
each rank first determines which CPU numbers are occupied by worker threads and saves them in a CPU mask.
The CPU mask is reduced on each compute node by splitting the MPI communicator with \texttt{MPI\_COMM\_TYPE\_SHARED}.
We then use libnuma to compute the list of NUMA domains covered by the worker threads.
Finally, we pin the communication and I/O threads to all free logical cores that are also in the list of used NUMA domains.
We thus ensure that pinning of pthreads is NUMA aware and does not interfere with the worker threads.
Note that throughout this study, we use simultaneous multithreading, i.e.\ two threads per physical core, as this results in better performance for our code.

\subsection{Load balancing}
SeisSol's static load balancing relies on graph-based partitioning.
The dual graph of a tetrahedral mesh is built by defining a vertex for each element and an edge for any two elements that share a face.
Vertex and edge weights model computation and communication cost, respectively.
SeisSol uses ParMETIS to solve the resulting graph partitioning problem~\cite{schloegel2002parallel}.
Vertex and edge weights are crucial for SeisSol's local time stepping scheme, as the element's update-rate is directly proportional to the element's computation cost~\cite{breuer_petascale_2016}.

New challenges raised by the fully-coupled model include:
\begin{itemize}
\item Our friction law requires the solution of a non-linear system of equations, cf.\ \cref{eq:bc-dynrup}.
    The number of Newton iterations varies over time such that the computational load is dynamic.
    \item The gravitational boundary requires additional operations due to the integration of the displacement in time.
    As discussed in \cref{sec:boundary-conditions}, we have to solve an ODE on the faces.
    Each Runge-Kutta stage requires an evaluation and extrapolation of the numerical solution.
\end{itemize}
The resulting variable performance, however, can only be addressed by assigning static (vertex) weights for partitioning.
To determine these weights,
we count for each vertex the dynamic rupture faces ($n_\text{DR}$) and the number of gravitational boundary faces ($n_G$) of the corresponding element.
The total weight is then given by
\begin{equation}
    2^{c_\text{max}-c_v} (w_\text{base} + w_\text{DR} n_\text{DR} + w_{G} n_{G}).
\end{equation}
The update rate of the element is reflected by $2^{c_\text{max}-c_v}$, where
$c_v$ is the cluster number of vertex $v$ and $c_\text{max}$ is the number of the largest time cluster~\cite{breuer_petascale_2016}.
As vertex weights are integers in ParMETIS, we assign a base weight of $w_\text{base}=100$ and determine $w_\text{DR}$ and $w_{G}$ relative to $w_\text{base}$.

We varied the weights $w_{DR}$ and $w_{G}$ in the range between 50 and 500 and ran a production simulation for \SI{0.2}{\s}.
For $w_{G}$, we found that the performance generally increases with weight, indicating that a weight in the range of 300--500
is appropriate.
For $w_{DR}$, a clear trend is not apparent.
Moreover, we face the following conceptual problems for $w_{DR}$:
First, the short simulated time for such test runs might not be indicative for a whole production run as the load is dynamic and expected to increase after earthquake nucleation.
Second, when a dynamic rupture face is adjacent to two partitions, the slip-rate on a fault face is computed twice.
That is, the load depends not only on the weights but on the partition itself.
Improving the dynamic rupture load balancing hence likely requires dynamic load balancing techniques.
We therefore set weights heuristically to $w_{DR}=200$ and $w_{G}=300$ in this work, which leads to an adequate parallel efficiency as shown in \cref{sec:results}.

Petascale supercomputers can suffer from strong performance fluctuations and few nodes may be substantially slower than others.
Thus, treating supercomputers as homogeneous can lead to a slowdown of the entire simulation~\cite{McCalpin:sc18,Wylie2020}.
Hence, we assume that their performance is inherently heterogeneous.
For each simulation run, we perform a small benchmark of our computational kernels prior to partitioning.
The inverse of the measured computational time for the small benchmark is taken as node weight.
Normalized weights are then fed into ParMETIS (\texttt{tpwgts}).
The performance impact of node weights is discussed in \cref{sec:results-palu,sec:results-scaling}.

\section{Results}\label{sec:results}
%We have discussed the derivation and implementation of our fully-coupled scheme in the previous section.
%% MB: no need to talk about the past ...
%In this section, we discuss the validation of our results.
Our fully-coupled elastic-acoustic model establishes a novel method for solving an entirely new class of earthquake-tsunami problems well beyond established earthquake and shallow water tsunami solvers. 
In particular, our model is not just an alternative way to compute the initial ocean displacement or even the tsunami wavefield. 
While the tsunami solution from 3D fully coupled modeling is very similar to that of the more traditional one-way linked modeling under certain simplifying conditions (cf.~\cref{sec:results-ascete}), the fully coupled model always computes the full wavefield that includes seismic and acoustic waves in addition to tsunami waves and allows non-linear coupling to earthquake physics.
In important parts of the parameter space, a fully coupled solution is essential since (i) the approximations that form the basis of the shallow water equations are invalid, (ii) the shallow water equations simply cannot capture seismic and acoustic waves that can be the dominant features in seismograms and pressure data recorded by offshore instruments over the source region and (iii) the complexity of the tsunami, acoustic and seismic ocean response phases and the influence of geometry, geologic structure and earthquake dynamics requires coupled solutions to fully understand tsunami generation processes \cite{Wirp2021,Elbanna2021}. %\todo[inline]{(have to add a point on tsunami generation dynamics)...} 

We therefore first validate our 3D fully-coupled Earth and ocean model against one-way linked earthquake-shallow-water modeling in a setup where the latter approximation is valid. We then present results of an extreme-scale application of the 2018 Sulawesi earthquake and tsunami, which showcases the added complexity of results obtained via the fully-coupled model.
We present strong scaling and performance analysis of the end-to-end application scenario.
We use maximum polynomial degree $N=5$ and $C(N) = 0.35 \left(2N+1\right)^{-1}$  for all simulations.
The performance and scaling tests and the production runs were executed on three different petascale systems:
\begin{description}
\item[\shaheen] 6174 nodes with dual socket Intel Xeon E5-2698 v3. Nodes are connected with an Aries interconnect and Dragonfly topology.
\item[\supermuc] 6336 nodes with dual socket Intel Skylake
Xeon Platinum 8174 (24 cores each).
The nodes are organized in 8 islands.
A fat tree OmniPath network is used, between each island the connection is pruned (pruning factor 1:4)
\item[\mahti] 1404 nodes with dual socket AMD Rome 7H12 (64 cores each).
Nodes are connected by a Mellanox HDR InfiniBand interconnect using a Dragonfly+ topology.
\end{description}

\subsection{Earthquake-Tsunami benchmark scenario}\label{sec:results-ascete}
\begin{figure*}[ht]
    \centering
    \includegraphics[width=0.8\textwidth]{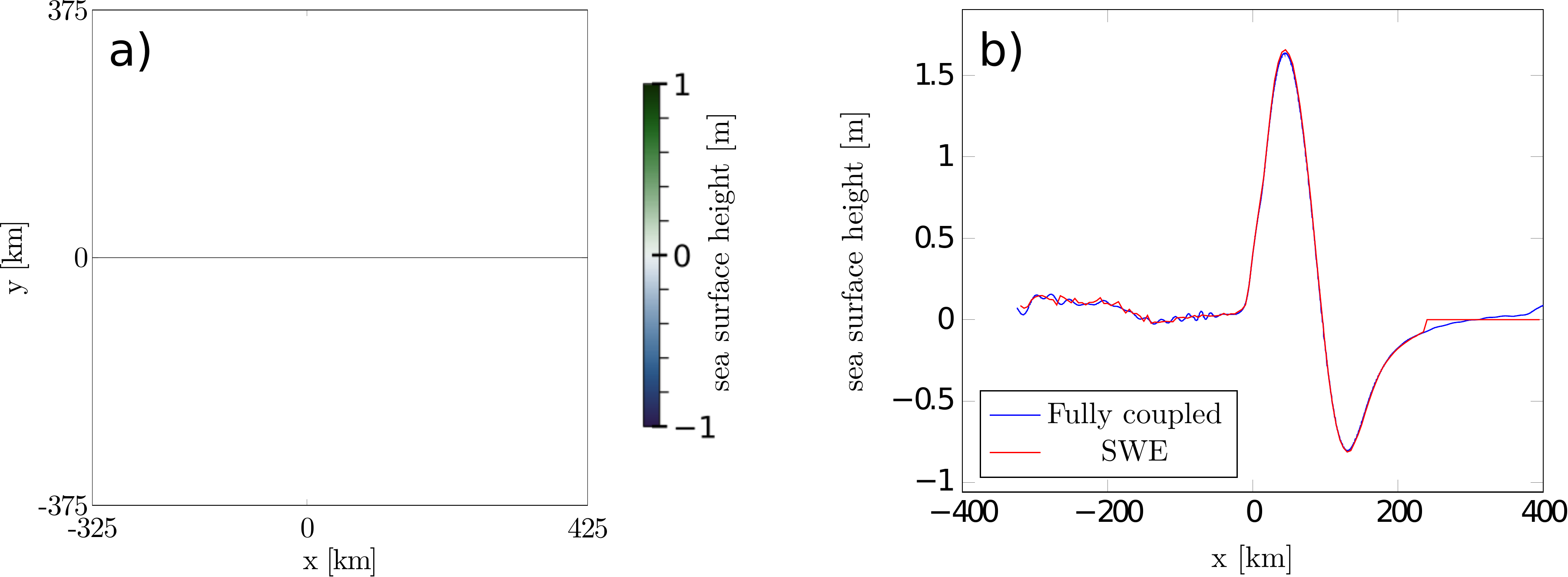}
    \caption{
   Sea surface height in a 3D megathrust earthquake-tsunami benchmark (\enquote{Scenario A} of \cite{madden_linked_2020}). (a) Snapshot of the sea surface height (vertical displacement $\eta$) of the fully coupled model at t=120\,s.
    %The black line indicates the location of the cross-section shown in (b). 
    (b) Comparison of the fully coupled model to a one-way linked model along a cross-section at y=0\,km (indicated in (a) as black line) 
    through the epicenter at t=120\,s.
    Blue: fully coupled sea surface height, including ocean response.
    Red: one-way linked model with a non-linear shallow water solver using the dynamic seafloor displacement as forcing term and including a linearly sloping beach at x=240\,km.
   Both models use the same dynamic rupture earthquake source performed with the same polynomial order 5.
   %Figure shows results at $t =  \SI{120}{\s}$.
    }
    \label{fig:results-ascete}
\end{figure*}
Verification of the correctness of the acoustic coupling and interface condition implementation has been demonstrated in preliminary convergence analyses with respect to analytic solutions~\cite{abrahams_verification_2019,krenz_elastic-acoustic_2019}. 
In the following, we compare the fully-coupled model with linked modeling for the \enquote{Scenario A} dynamic megathrust earthquake benchmark of~\citet{madden_linked_2020}.
%, to demonstrate that the fully-coupled model captures correct propagation of the tsunami (due to the linearized gravity boundary condition). 
%We here begin our investigation of rupture dynamics and the generation and propagation of tsunami and other waves with the \enquote{Scenario A} dynamic megathrust earthquake benchmark of~\citet{madden_linked_2020}. 
%\todo[inline]{requires/d rewording (first attempt done ...)}
This setup includes an idealized 3D megathrust dynamic rupture earthquake source on a planar fault, tsunami generation and propagation in a basin with flat bathymetry and inundation on a linearly sloping beach.
%The fault is \SI{200}{\kilo \meter} wide and extends from the surface to \SI{35}{\kilo \meter} depth at a 16$^\circ$ dip. 
The benchmark is run with both our fully coupled model and a one-way linked 3D-2D modeling approach using a shallow water solver for tsunami propagation (as in~\cite{madden_linked_2020}). While we anticipate some level of agreement between the two models, differences in ocean response are expected as a consequence of the approximations employed in the linked model. Specifically, we expect the excitation of ocean acoustic modes in the fully coupled simulation within the deep water (2~km depth) by the high-frequency seismic waves resolved in the earthquake model~\cite{abrahams2021}. Further differences are expected since the fully coupled model does not include the sloping beach.
%This scenario features an idealized subduction zone, with flat bathymetry and a curved fault.   
%Due to the limited complexity of the scenario (especially neglecting complicated bathymetry), we expect that a two-step 3D-2D linked simulation approach (as in \cite{madden_linked_2020}) and our fully-coupled elastic-acoustic approach should lead to very similar results~\cite{abrahams2021}.    

%mesh
In \citet{madden_linked_2020}'s Scenario A, the fault is \SI{200}{\kilo \meter} wide and extends from the surface to \SI{35}{\kilo \meter} depth at a 16$^\circ$ dip. 
We discretize the fully-coupled scenario with tetrahedral edge length of \SI{400}{\m} at the fault and gradually coarsen the mesh away from the source.
%The mesh is gradually coarsened away from the source, %reaching a maximum edge length of \SI{100}{\kilo \meter}.
We add a \SI{2}{\km} water layer atop the Earth model. % to the dynamic rupture earthquake model with a height of \SI{2}{\km} (constrained by the parameters of the comparison tsunami model in \cite{madden_linked_2020}) and a coarsest horizontal resolution of $\SI{2}{\km}$.
Since %As the fault is meshed with finer resolution and because 
we use conforming meshes, the resolution of the water layer is higher near the earthquake source (locally refined to conform with a \SI{66}{\m} edge length at the megathrust fault).
Adding the water layer increases the mesh size from roughly 16 million (for the earthquake model) to 29.5 million elements for the fully coupled model.

%b.c.
On-fault earthquake dynamics are governed by a linear-slip weakening friction law~\cite{Andrews1976}, which is computationally less demanding than rate-and-state friction employed in the complex Palu, Sulawesi application (\cref{sec:results-palu}). Higher fault strength near the seafloor smoothly stops the rupture as it approaches the Earth surface. The earthquake magnitude is $M_w=$~8.5 and the rupture propagates at  \SI{3.5}{\kilo\metre/\second} on average.
%At the water surface, we apply the gravitational free surface boundary condition.
%structure
The solid has homogeneous elastic properties representative of oceanic crust in a subduction zone~\cite{Stephenson2017}: 
%c_p = ((8.5289429e10+2*6.7526085e10)/3775)**0.5  = 7639.936877794786 
%c_s = (6.7526085e10/3775.0)**0.5 =  4229.385846167454
 p-wave speed $c_p = \SI{7639.9}{\m/\s}$, shear wave speed $c_s = \SI{4229.4}{\m/\s}$ and density $\rho = \SI{3775}{\kilogram/\meter^3}$.
 %
 % c_p = sqrt(2.25e+9/1000)
The ocean has an acoustic wave speed of $c_p = \SI{1500}{\m/\s}$ and density $\rho = \SI{1000}{\kilogram/\meter^3}$.

We compare the fully-coupled simulation to the 
one-way linked reference solution in \cref{fig:results-ascete}.
%Both simulations use the same SeisSol dynamic rupture earthquake simulation with polynomial order 5.
%The displacement at the ocean floor is recorded.
In the one-way linked approach, the seafloor displacement recorded on the unstructured mesh of the earthquake model is bilinearly interpolated to an intermediate uniform Cartesian mesh with a resolution of $\SI{1000}{\meter}$, which is subsequently used as a time-dependent source in the hydrostatic non-linear shallow water tsunami model.
The time-dependent forcing term in the shallow water mass balance is here taken as the unfiltered seafloor displacement field.
Then the second-order Runge-Kutta DG non-linear shallow water solver \samoaflash{} is used to perform the tsunami simulation using dynamically adaptive mesh refinement (see \citet{madden_linked_2020} for details). 
%The \samoaflash{} tsunami model has been implemented within the \samoa\ framework, which is publicly available as open source software at \url{https://gitlab.lrz.de/samoaterm in the shallow water mass balance
%Here, the recorded seafloor displacement is directly used as a time-dependent forcing term in the shallow water mass balance.
%We did not apply any filters on the displacement field, in distinction to Scenario A in \cite{madden_linked_2020}.

The sea surface height from our fully coupled solution matches the one-way linked approach at the low frequencies characterizing the tsunami response (\cref{fig:results-ascete}, b).
As expected, differences appear at $x>\SI{240}{\kilo \meter}$, at the right-hand side of the domain, where the reference simulation includes a beach which is not contained in our fully coupled model. The short wavelength and high frequency oscillations (periods $<\SI{5.3}{\s}$) trailing the leading seismic wavefronts (\cref{fig:results-ascete}, a) are reverberating acoustic wave modes in the ocean, which are captured only in our fully coupled model.

%Verification of the correctness of the acoustic coupling and interface condition implementation is demonstrated in preliminary convergence analysis with respect to analytic solutions~\cite{abrahams_verification_2019,krenz_elastic-acoustic_2019}. 

\subsection{Dynamics of the 2018 Palu, Sulawesi tsunami generation}\label{sec:results-palu}
Next, we present the first fully-coupled 3D simulation of a real event: the 2018 magnitude 7.5 strike-slip supershear Sulawesi earthquake and associated local tsunami in Palu Bay (\cref{fig:teaser}). %
%and compare to the one-wave linked earthquake–tsunami model of \citet{ulrich_coupled_2019}. 
The dynamic rupture earthquake source model features sustained supershear rupture propagation across a multi-segment fault model (\cref{fig:teaser}, c) and matches key observed earthquake characteristics, including the moment magnitude, rupture duration, fault plane solution, teleseismic waveforms and inferred horizontal ground displacements from observations of optical and radar satellites~\cite{ulrich_coupled_2019}. It incorporates a fast velocity weakening rate-and-state friction law~\cite{harris_suite_2018,pelties_gmd_2014} which is motivated by laboratory rock physics experiments. The time-dependent 3D seafloor displacements produce a mean vertical uplift of 1.5~m underneath Palu Bay. This sources a tsunami with wave amplitudes and periods that match those measured and inundation that reproduces observations from field surveys.

The fully coupled model includes a water layer of variable depth atop BATNAS bathymetry and topography data of horizontal resolution of 6 arc seconds ($\approx \SI{190}{\m}$, \citep{DEMNAS2018}).
Due to the bathtub-like, steep geometry of Palu Bay and the shallow water depth (on average \SI{600}{\meter}), the locally adaptive unstructured tetrahedral mesh is crucial for rendering this model feasible.
Adaptive local refinement as fine as \SI{200}{\m} element edge length ensures accurate resolution of rupture dynamics across the complex fault system \citep{Wollherr2018} -- we observe high frequency seismic wave radiation of up to 10 Hz close to the faults within Palu Bay.
We set a maximum global element size of $\SI{5000}{\m}$ and refine the resolution in the water layer and in our region of interest:
a cuboid centered at $x=\SI{5}{\kilo \meter}, y = \SI{0}{\kilo \meter}, z=\SI{-8}{\kilo \meter}$ with extent $ x=\SI{70}{\kilo \meter}, y = \SI{180}{\kilo \meter}, z=\SI{34}{\kilo \meter}$.

We present \enquote{production runs} using the following two meshes.
The respective numbers of degrees of freedom using a polynomial order of 5 are given in parentheses.
\begin{description}
\item[M] Our medium-sized mesh with $\approx$89 million elements ($\approx$ 46 billion degrees of freedom) has a water layer resolution of $\SI{100}{\meter}$ and resolves seismic wave propagation with elements of a maximum size of \SI{1000}{\meter}.
\item[L] Our large mesh has $\approx$518 million elements ($\approx$ 261 billion degrees of freedom).
Its resolution in both the water layer with $\SI{50}{\meter}$ and in the seismic wave refinement zone with \SI{500}{\m} is twice as fine as in the \textbf{M} mesh.
\end{description}
In the \textbf{L} mesh, 453.7 million grid cells are used to discretize the ocean (acoustic layer), which increases the total mesh size by a factor of 8. 
This makes the simulation of the \textbf{L} mesh almost twice as large in terms of degrees of freedom as the so-far largest reported SeisSol simulation~\cite{uphoff_extreme_2017}.

We run the \textbf{M} mesh for \SI{100}{\s} which is long enough for the localized tsunami to interact with the coastline of the bay.
The higher resolution \textbf{L} model is run for $\SI{30}{\s}$ during which we capture details of the entire earthquake rupture process and the generation and interaction of both acoustic waves and tsunami.
Within the water layer, we prescribe a vertical and horizontal spatial discretization of maximum \SI{50}{\m} element edge length in the \textbf{L} mesh, resolving a maximum frequency of at least \SI{15}{\hertz} of the acoustic wave field (ensuring that 2 elements of polynomial order 5 space-time accuracy sample one wavelength propagating at an approximate speed of \SI{1483}{\meter / \second}~\cite{kaser_quantitative_2008}).
With this setup, we measure wave excitation of up to \SI{30}{\hertz} in the Fourier spectra of the recorded acoustic velocity time series which we attribute to the highly variable water depth.

The high-resolution simulation using the \textbf{L} mesh is illustrated in \cref{fig:teaser}.
%velocity
The velocity field snapshot in \cref{fig:teaser} (a) reveals a very high degree of complexity at \SI{15}{\s} simulation time. A sharp shear Mach front is stacked up due to the fast moving dynamic rupture source. Seismic waves within Earth, acoustic waves within the ocean (expected at periods shorter than $\approx$\SI{1.6}{\s}) and wave conversions superimpose. 
 At this point in time during the simulation, the unilaterally propagating southward rupture arrives at the end of Palu Bay (\cref{fig:teaser}, b).
The vertical displacement field of our high-resolution fully-coupled model in \cref{fig:teaser} (c) shows that seismic waves (including the sharply imprinting supershear Mach front) result in transient motions of the sea surface and affect the ocean response but do not appear to contribute to tsunami generation.

\begin{figure}
    \centering
    \includegraphics[width=0.8\columnwidth]{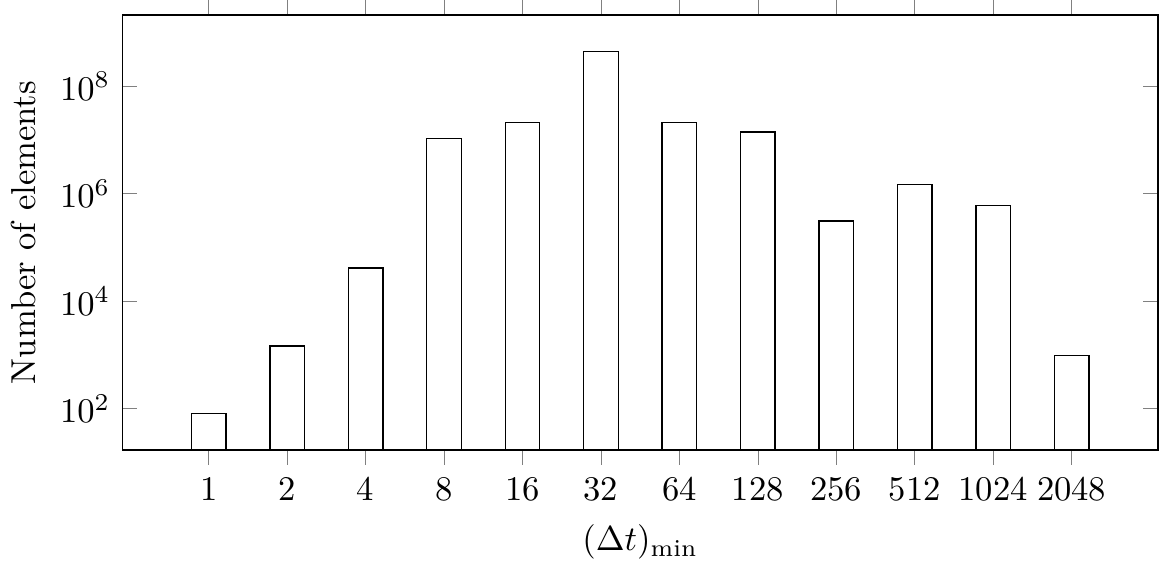}
    \caption{Distribution of elements to the local time-stepping clusters for mesh \textbf{L}.
    The x-axis gives the timestep size of the clusters in terms of the minimum timestep size $(\Delta t)_\text{min}$.
    Note that the y-axis is scaled logarithmically.}
    \label{fig:clustering-palu}
\end{figure}

Time to solution, specifically of the large production run, benefits greatly from local time-stepping:
The chosen clustering (\cref{fig:clustering-palu}) reduces the total number of required element updates by a factor of $\approx$30.
More than 86\% of all elements fall within the cluster with timestep $32 (\Delta t)_\text{min}$.
Running with global time-stepping is therefore not feasible and we thus do not report respective performance or simulation results. 
%This is why we only report results obtained with LTS in this paper, as global time-stepping is not a realistic solution strategy.
%
We ran the \textbf{L} mesh on 3072 nodes of \supermuc{} and achieved 3.14~\petaflops\ on average, i.e., a bit more than 1~\teraflops\ per node.
The run took 5 hours and 30 minutes to completion (time reported by Slurm), hence running it for the same time as the \textbf{M} mesh would take 18 hours and 15 minutes, which is computationally feasible.
We ran the same setup on 6144 nodes of \shaheen{}, reaching an estimated performance%
\footnote{The simulation timed out due to the allocated job duration after reaching $\approx \SI{16}{\s}$.
We estimate performance by interpolation from the FLOP count of the \supermuc{} run and from the measured execution time.}
of 2.3~\petaflops.
These simulations (and performance numbers) include receiver output (every \SI{0.01}{\s}) and free surface output (including the elastic-acoustic interface, every \SI{0.1}{\s}).
The merits of assigning node weights become clear for these large-scale runs.
For the \textbf{L} run on \supermuc{}, we see node weights of $\approx 4.54 \pm 0.087$ ($\text{mean} \pm \text{standard deviation}$) with a minimum weight of just 2.74.
That means the slowest node had a performance of only 60.4\% of the average node.
On \shaheen{}, performance variability was less severe, with node weights of  $3.34 \pm 0.023$ with a minimum of 3.19.

\begin{figure}
    \includegraphics[width=\linewidth]{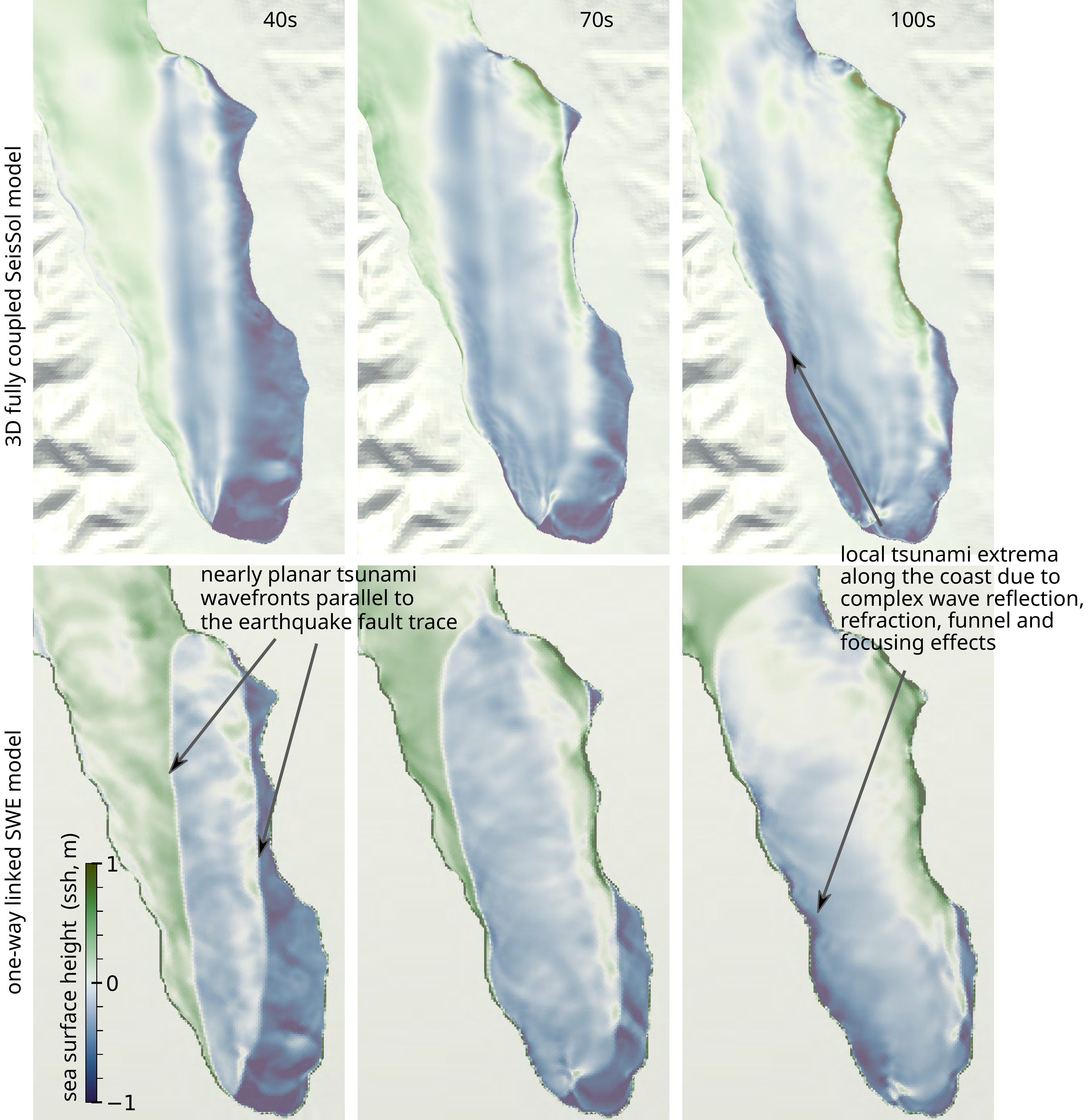}
    \caption{Snapshots of vertical displacement of the ocean surface within Palu Bay in the fully-coupled Sulawesi earthquake-tsunami simulation using mesh \textbf{M} (upper row) compared to a one-way linked 2D shallow water tsunami simulation (lower row). While both models capture the overall dynamics of the dynamic earthquake-ocean response we observe differences in the sharpness of wave fronts (see text) in addition to the excitation of acoustic waves as was observed in the deep ocean benchmark (fast seismic and acoustic waves are not shown here but visible in \cref{fig:teaser,fig:results-ascete}). Fully-coupled 3D simulations open the possibility to explore the effects of variable depth ocean non-hydrostatic response, compressibility and acoustic wave generation across complex bathymetry.
    %Palu bay is shown, with surrounding area.
    }
    \label{fig:palu-displacement}
\end{figure}

\Cref{fig:palu-displacement} shows a series of snapshots in time of the vertical displacement fields evolving during the \textbf{M} mesh earthquake-tsunami simulation in the upper row.
The lower row shows the same snapshots generated with a one-way linked method that uses a non-linear shallow water solver.
%There is a large amount of subsidence (southeast) and uplift (northwest) from the static displacement of the solid.
The ocean experiences subsidence (southeast of the fault) and uplift (northwest of the fault) from the static earthquake displacements (\cref{fig:teaser}, d) as modulated by topography and bathymetry. 
As the rupture propagates much faster than the tsunami wave speed, the tsunami is sourced with nearly planar wavefronts parallel to the fault trace (\cref{fig:palu-displacement}, \SI{40}{\s}). Deviations are most pronounced where the fault enters and exits the bay and are primarily caused by the slower wave speed in the shallow water along the edge of the bay. 
%
%The one-way linking approach produces a tsunami with much sharper wavefronts, which we speculate may be caused by the hydrostatic approximation used in the shallow water solver and aided by the very complex bathymetry in Palu Bay.
%At \SI{40}{\s}, the tsunami propagating away from the sharp step-wise seafloor displacement at the earthquake fault trace (cf. \cref{fig:teaser}) towards the coast. 
%In the following, the earthquake subsidence and uplift pattern is perturbed by the fairly sharp tsunami front, while 
Such clear patterns are not visible anymore at \SI{100}{\s}.
The local extrema in vertical displacement along the coast illustrates the complex wave reflections and refractions within the bay caused by shallow bathymetry as well as funnel and focusing effects.
We observe continuous reflections of waves from the coast, for example, on the lower-left corner.% (see also the velocity vector field in~\cref{fig:palu-streamlines})
%which are also apparent in \cref{fig:palu-streamlines}, which visualizes the velocity vector field.

% adding Eric's text here. I find the physical part too strong, I phrase it slightly more carefully
%The upper and lower rows compare the sea surface uplift for the fully coupled model with that from a two-step method that uses a nonlinear shallow water solver. 
While most wavefield features are quite similar, as are predicted wave heights, there are also clear differences. Both simulations are performed with the same bathymetry and coastlines and similar element sizes, suggesting that differences may be physical and may arise from the inclusion of additional processes, such as compressibility, in the fully coupled model. The one-way linking approach produces a tsunami with much sharper wavefronts, which we speculate may be caused by the hydrostatic approximation used in the shallow water solver and aided by the very complex bathymetry in Palu Bay. The wavefield is notably smoother in the fully coupled model, perhaps due to non-hydrostatic effects that filter short-wavelength features in the transfer function between seafloor and sea surface motions~\cite{kajiura1963} or due to conversion between surface gravity, acoustic, and seismic wave modes. Also, nonlinear effects may become important during wave interaction with the coastline. %(\cref{fig:palu-streamlines}.
%Another possibility is the inclusion of nonlinearity in the shallow water solver, which might become important during wave interaction with the coastline.
Fully-coupled 3D simulations allow us to explore these hypotheses in great detail and, together with extended numerical analysis, to better understand when one-way linked approaches are justified and when fully coupled modeling is required.

%\begin{figure}
%    \centering
%    %\includegraphics[width=0.6\columnwidth]{steamlines_edited}
%    \includegraphics[width=0.9\columnwidth]{steamlines_edited_rotated}
%    \caption{Vertical displacement at 100~s using mesh \textbf{M}. % SI breaks formatting here!
%    The white and black lines are velocity field streamlines which are instantaneously tangent to the vertical velocity vectors of the ocean within Palu Bay.
%    At this point in time, streamlines originate upwards and downwards from the submerged fault trace along the Bay.
%    }
%    \label{fig:palu-streamlines}
%\end{figure}

%Alice: I moved this text block describing Fig. 1 up

\subsection{Strong scaling}\label{sec:results-scaling}
The SeisSol software has been shown to scale well for dynamic rupture earthquake simulations on various petascale machines~\cite{heinecke_petascale_2014,uphoff_extreme_2017}.
However, the extensions to the presented coupled elastic-acoustic model impose several new challenges for load balancing and on the cluster-oriented local time-stepping algorithm (due to the smaller timesteps in the acoustic layer and the resulting different distribution of elements to timestep clusters).
The main challenge for coupled elastic-acoustic computation is resolving the high frequency range involved and the low wave speeds, e.g., in near-surface sediments, requiring locally high spatial discretization~\cite{Rosenkrantz2019}.
We therefore present scalability results of the Palu, Sulawesi earthquake-tsunami application scenario on \mahti\ and \supermuc. 
While SeisSol has previously been optimized towards the use of a single MPI rank per compute node, we observe (see \cref{sec:node-level-performance}) stronger influence of NUMA domains on node-level performance on recent architectures. 
We therefore experiment with different numbers of MPI ranks in the scalability tests and particularly include \mahti\ as an AMD-based machine, whose CPUs feature 4 NUMA domains.
\begin{figure}
    \centering%
    \begin{subfigure}[t]{0.77\columnwidth}
    \caption{Parallel efficiency on \mahti{}:
    \label{fig:scaling-study-mahti}}
    \includegraphics[width=\textwidth]{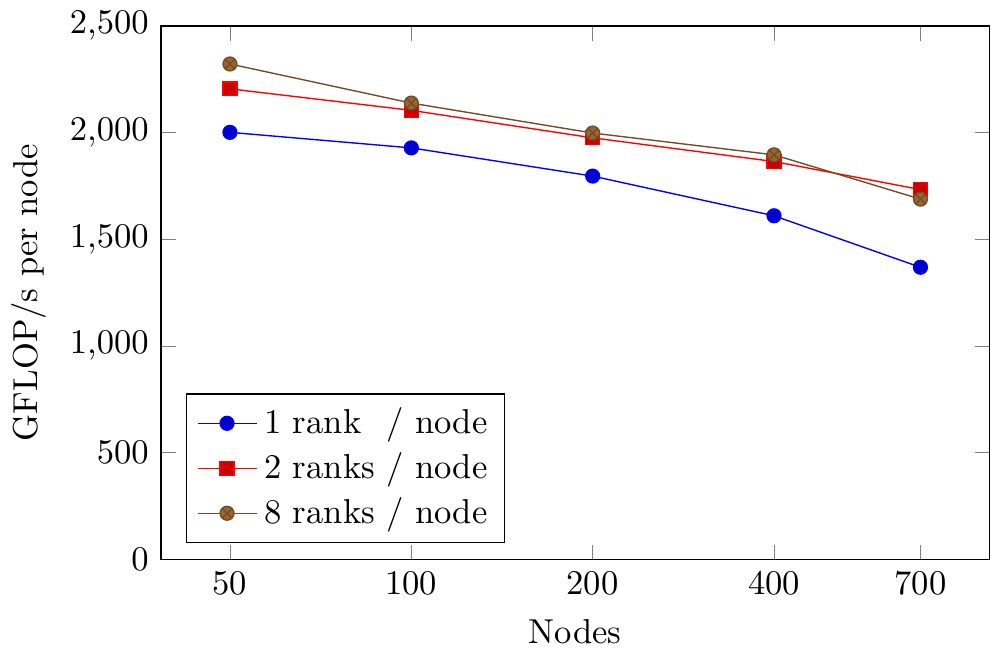}
    \end{subfigure}
    \begin{subfigure}[t]{0.77\columnwidth}
    \caption{Parallel efficiency on \supermuc{}:
    \label{fig:scaling-study-ng}}
    \medskip
    \includegraphics[width=\textwidth]{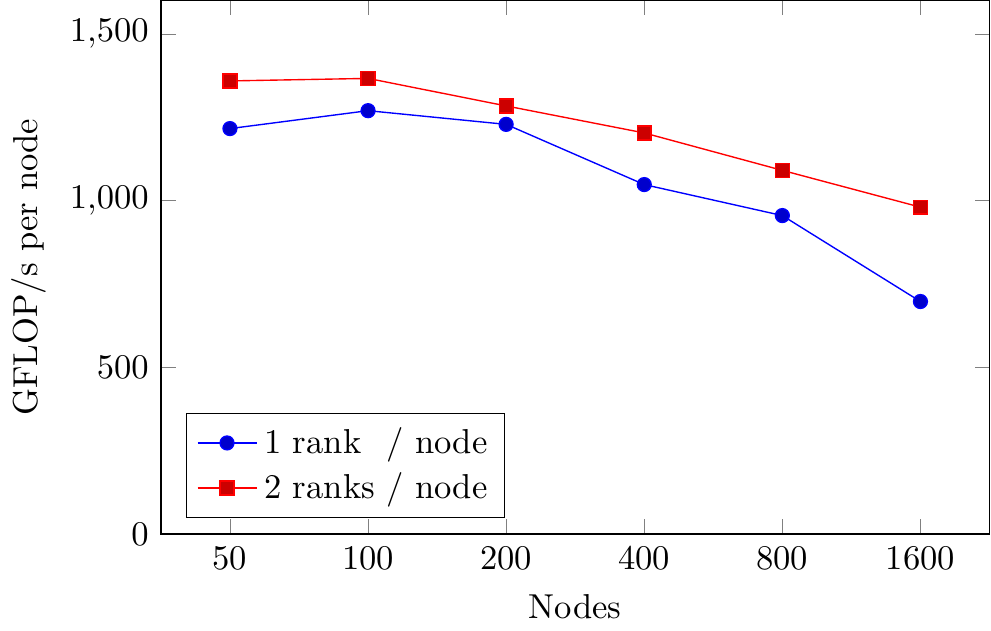}
    \end{subfigure}%
    \caption{Strong scaling for mesh \textbf{M} on \mahti\ and \supermuc\ (equivalent to parallel efficiency).}
    \label{fig:scaling-study}
\end{figure}
As described in \cref{sec:hpc}, we expect that a larger number of MPI ranks per node should have a beneficial effect, due to NUMA effects. 
To evaluate the optimal number of nodes per rank, we ran our scaling study (\cref{fig:scaling-study-mahti}) for 1 MPI process per node, per socket (2 ranks per node) and per NUMA domain (8 ranks per node).
The situation on \supermuc{} is similar but less severe, with two NUMA domains per node, stemming from the dual socket configuration.
Thus, we ran the scaling tests (\cref{fig:scaling-study-ng}) to compare 1 process per node with 1 process per socket.

We ran all scaling studies until \SI{0.03}{\s} which includes several LTS cycles.
In contrast to our production runs, we deactivated output for our scaling study.
We report the maximal performance over multiple runs to avoid performance fluctuations.
Note that we sacrifice one physical thread of each MPI process for efficient communication (MPI progression).
Despite this, we achieved best results using one process per NUMA domain.

Using this setup, we achieved 1359 \gigaflops{} per node with 50 nodes on NG and 981 \gigaflops{} on 1600 nodes.
This translates to a parallel efficiency of $\approx$72\%.
On Mahti, we achieved 2322 \gigaflops{} on 50 nodes and 1689 \gigaflops{} on 700 nodes, resulting in a parallel efficiency of $\approx$73\%.
We achieved up to $\approx$1.57 \petaflops{} on \supermuc{} and up to $\approx$1.18 \petaflops{} on \mahti{} in total.
We also ran a scaling study on the \textbf{L} mesh on \supermuc{}. Using 2 nodes per rank, we achieved a performance of 1298 \gigaflops{} on 768 nodes and of 996 \gigaflops{} on 3072 nodes. This corresponds to a parallel efficiency of $\approx 76.8\%$.
Furthermore, the \textbf{L} mesh ($\approx$518 million elements) on 3072 nodes of \supermuc{} corresponds to a workload of $\approx$$1.68 \cdot 10^5$ elements per node, which is equivalent to using roughly 500 nodes for the \textbf{M} mesh. 
As the latter falls within the region of good parallel efficiency in \cref{fig:scaling-study}, we expect that the \textbf{L} mesh scenario will also scale well on the full \supermuc{} machine.

Finally, we used the strong scaling setup to evaluate the effect of our node weights.
When we deactivated this feature, we got a maximum performance (on 700 nodes of \mahti{}) of 1412 \gigaflops{}, i.e.\ 84\% of the performance with node weights.

\section{Conclusion and outlook}
We developed a 3D fully-coupled model for earthquake-tsunami interaction that includes tsunami propagation, and described our novel discretization and implementation in the software SeisSol.
Our numerical scheme is high order in both space and time and allows the simulation of multi-physics scenarios that include frictional slip on faults and wave propagation in both elastic and acoustic layers with gravity.
%
%Our contributions to HPC are the following:
%we achieve excellent single node performance on the AMD Rome architecture (up to 63\% for predictor step),
%we described our heuristic for multi-physics load-balancing and we explained, how we achieve a fairer load-balancing by treating supercomputers as inherently heterogeneous.
%
We showed results for an earthquake-tsunami interaction benchmark that demonstrate that our method compares well to solutions obtained with \enquote{one-way} linking with a shallow water model under certain conditions.

We presented a novel, fully-coupled setup for the 2018 Palu, Sulawesi earthquake and tsunami that captures the dynamics of the entire tsunami-genesis in a single simulation. There are notable differences in the fully coupled and one-way linked models for the Palu event, raising questions regarding the validity of approximations used in standard earthquake-tsunami coupling workflows. Only with a fully coupled solution can these approximations be explored in the complex geometries and material structures representative of real events.
%Our production runs achieve petascale performance on both \supermuc{} and \shaheen.
%Our new simulation capabilities open the possibility to fundamentally further our understanding of earthquake-tsunami interaction in its full complexity, which includes identifying when one-way linked approaches are justified and when fully coupled modeling is required.

Our production runs achieved petascale performance on both \supermuc{} and \shaheen.
Our solver scales up to 1600 nodes on \supermuc{} and up to 700 nodes on \mahti{} for our \textbf{M} production scenario, achieving excellent single node performance (up to 63\% for the predictor step on the AMD Rome architecture).
We described our heuristic for multi-physics load-balancing and we explained, how we achieve a fairer load-balancing by treating supercomputers as inherently heterogeneous.
We discussed limits of the static load balancing, which becomes increasingly difficult when combining local time-stepping and multi-physics and which cannot deal with dynamic load, as posed by the dynamic earthquake rupture friction law.
In future work, we want to explore a task-based execution of clustered local time-stepping which enables the use of remote task off-loading for dynamic load balancing~\cite{klinkenberg2019}.
%Finally, we want to run the acoustics part without embedding it into the elastic wave equation in the future.
%This requires complex restructuring of the code but would also ease additional multi-physics coupling in the future.

Our work opens up new avenues for research into earthquake-tsunami interaction.
Not only is it now possible to capture the entire dynamics of this process in a unified 3D model; it is also efficient.
This permits fully-coupled simulations even for large scenarios, instead of relying on approximate 3D-2D linked methods.
Our new simulation capabilities open the possibility to fundamentally improve our understanding of earthquake-tsunami interaction in its full complexity, which includes identifying when 3D-2D one-way linked models are justified and when fully coupled modeling is required.
Also, our implementation will be instrumental to further elastic-acoustic coupling applications, such as holistic modeling of earthquake-generated acoustic waves~\cite{komatitsch2011} in complex geological settings or seismic waves originating from lava lake sloshing~\cite{liang2020lava}.

%%
%% The acknowledgments section is defined using the "acks" environment
%% (and NOT an unnumbered section). This ensures the proper
%% identification of the section in the article metadata, and the
%% consistent spelling of the heading.
\begin{acks}
This project has received funding from the European Union's Horizon 2020 Research and Innovation Programme under grant agreement No. 823844 (ChEESE -- Centre of Excellence in Solid Earth).
% thank colleagues from ChEESE here (e.g., general thanks for the collaboration?)
% MB: Would like to single out this project acknowledgement - have it first, then the compute resources, and then maybe additional projects, where necessary. 
%
Compute resources were provided 
by the Gauss Centre for Supercomputing e.V.\ (www.gauss-centre.eu) on SuperMUC-NG at the Leibniz Supercomputing Centre (www.lrz.de, project pn68fi),
by the CSC -- IT Center for Science, Finland (project 2003841, on Mahti) 
and by the Supercomputing Laboratory at King Abdullah University of Science \& Technology (KAUST, project k1488, on Shaheen-II) in Thuwal, Saudi Arabia.
% The authors gratefully acknowledge the Gauss Centre for Supercomputing e.V. (www.gauss-centre.eu) for funding this project by providing computing time on the GCS Supercomputer SuperMUC-NG at Leibniz Supercomputing Centre (www.lrz.de).
% The authors wish to acknowledge CSC – IT Center for Science, Finland, for computational resources.
% Projects are: pn68fi (NG), project 2003841 (Mahti) and k1488 (Shaheen).
We thank all colleagues at LRZ, CSC and KAUST for their excellent support. 
%in realising the simulations and scalability tests on SuperMUC-NG, Mahti and Shaheen-II. 
% Should we name certain persons individually? 
% For computer time, this research used the resources of the Supercomputing Laboratory at King Abdullah University of Science & Technology (KAUST) in Thuwal, Saudi Arabia
% The authors gratefully acknowledge the Gauss Centre for Supercomputing e.V. (www.gauss-centre.eu) for funding this project by providing computing time on the GCS Supercomputer SuperMUC-NG at Leibniz Supercomputing Centre (www.lrz.de).
% The authors wish to acknowledge CSC – IT Center for Science, Finland, for computational resources.
% Projects are: pn68fi (NG), project 2003841 (Mahti) and k1488 (Shaheen).
%
C.U., T.U. and A.-A.G. acknowledge additional support by the European Union’s Horizon 2020 Research and Innovation Programme under ERC StG TEAR, no. 852992, the German Research Foundation (DFG) (grants no. GA 2465/2-1, GA 2465/3-1) and by KAUST-CRG (grant no. ORS-2017-CRG6 3389.02).
L.S.A. was supported by National Science Foundation grant DGE-1656518.
\end{acks}

%%
%% The next two lines define the bibliography style to be used, and
%% the bibliography file.
\bibliographystyle{ACM-Reference-Format}
\bibliography{bibliography}

%%% -*-BibTeX-*-
%%% Do NOT edit. File created by BibTeX with style
%%% ACM-Reference-Format-Journals [18-Jan-2012].

\begin{thebibliography}{67}

%%% ====================================================================
%%% NOTE TO THE USER: you can override these defaults by providing
%%% customized versions of any of these macros before the \bibliography
%%% command.  Each of them MUST provide its own final punctuation,
%%% except for \shownote{}, \showDOI{}, and \showURL{}.  The latter two
%%% do not use final punctuation, in order to avoid confusing it with
%%% the Web address.
%%%
%%% To suppress output of a particular field, define its macro to expand
%%% to an empty string, or better, \unskip, like this:
%%%
%%% \newcommand{\showDOI}[1]{\unskip}   % LaTeX syntax
%%%
%%% \def \showDOI #1{\unskip}           % plain TeX syntax
%%%
%%% ====================================================================

\ifx \showCODEN    \undefined \def \showCODEN     #1{\unskip}     \fi
\ifx \showDOI      \undefined \def \showDOI       #1{#1}\fi
\ifx \showISBNx    \undefined \def \showISBNx     #1{\unskip}     \fi
\ifx \showISBNxiii \undefined \def \showISBNxiii  #1{\unskip}     \fi
\ifx \showISSN     \undefined \def \showISSN      #1{\unskip}     \fi
\ifx \showLCCN     \undefined \def \showLCCN      #1{\unskip}     \fi
\ifx \shownote     \undefined \def \shownote      #1{#1}          \fi
\ifx \showarticletitle \undefined \def \showarticletitle #1{#1}   \fi
\ifx \showURL      \undefined \def \showURL       {\relax}        \fi
% The following commands are used for tagged output and should be
% invisible to TeX
\providecommand\bibfield[2]{#2}
\providecommand\bibinfo[2]{#2}
\providecommand\natexlab[1]{#1}
\providecommand\showeprint[2][]{arXiv:#2}

\bibitem[\protect\citeauthoryear{Abrahams, Dunham, Krenz, Saito, and
  Gabriel}{Abrahams et~al\mbox{.}}{2021}]%
        {abrahams2021}
\bibfield{author}{\bibinfo{person}{Lauren~S. Abrahams},
  \bibinfo{person}{Eric~M. Dunham}, \bibinfo{person}{Lukas Krenz},
  \bibinfo{person}{Tatsuhiko Saito}, {and} \bibinfo{person}{Alice-Agnes
  Gabriel}.} \bibinfo{year}{2021}\natexlab{}.
\newblock \showarticletitle{Comparison of techniques for coupled earthquake and
  tsunami modeling}.
\newblock \bibinfo{journal}{\emph{Earth and Space Sci. Open Arch.}}
  (\bibinfo{date}{Feb.} \bibinfo{year}{2021}), \bibinfo{pages}{49}.
\newblock
\urldef\tempurl%
\url{https://doi.org/10.1002/essoar.10506178.3}
\showDOI{\tempurl}
\newblock
\shownote{AGU 2020 Fall Meeting.}


\bibitem[\protect\citeauthoryear{Abrahams, Krenz, Dunham, and Gabriel}{Abrahams
  et~al\mbox{.}}{2019}]%
        {abrahams_verification_2019}
\bibfield{author}{\bibinfo{person}{Lauren~S Abrahams}, \bibinfo{person}{Lukas
  Krenz}, \bibinfo{person}{Eric~M Dunham}, {and} \bibinfo{person}{Alice-Agnes
  Gabriel}.} \bibinfo{year}{2019}\natexlab{}.
\newblock \showarticletitle{Verification of a 3D fully-coupled earthquake and
  tsunami model}. In \bibinfo{booktitle}{\emph{AGU Fall Meet. Abstr.}},
  Vol.~\bibinfo{volume}{2019}. \bibinfo{pages}{NH43F--1000}.
\newblock
\urldef\tempurl%
\url{https://agu.confex.com/agu/fm19/meetingapp.cgi/Paper/547532}
\showURL{%
\tempurl}


\bibitem[\protect\citeauthoryear{Andrews}{Andrews}{1976}]%
        {Andrews1976}
\bibfield{author}{\bibinfo{person}{D.~J. Andrews}.}
  \bibinfo{year}{1976}\natexlab{}.
\newblock \showarticletitle{Rupture velocity of plane strain shear cracks}.
\newblock \bibinfo{journal}{\emph{J. Geophys. Res.}} \bibinfo{volume}{81},
  \bibinfo{number}{32} (\bibinfo{date}{nov} \bibinfo{year}{1976}),
  \bibinfo{pages}{5679--5687}.
\newblock
\urldef\tempurl%
\url{https://doi.org/10.1029/JB081i032p05679}
\showDOI{\tempurl}


\bibitem[\protect\citeauthoryear{Aniko~Wirp, Gabriel, Schmeller, Madden, van
  Zelst, Krenz, van Dinther, and Rannabauer}{Aniko~Wirp et~al\mbox{.}}{2021}]%
        {Wirp2021}
\bibfield{author}{\bibinfo{person}{Sara Aniko~Wirp},
  \bibinfo{person}{Alice-Agnes Gabriel}, \bibinfo{person}{Maximilian
  Schmeller}, \bibinfo{person}{Elizabeth~H. Madden}, \bibinfo{person}{Iris van
  Zelst}, \bibinfo{person}{Lukas Krenz}, \bibinfo{person}{Ylona van Dinther},
  {and} \bibinfo{person}{Leonhard Rannabauer}.}
  \bibinfo{year}{2021}\natexlab{}.
\newblock \showarticletitle{3D Linked Subduction, Dynamic Rupture, Tsunami, and
  Inundation Modeling: Dynamic Effects of Supershear and Tsunami Earthquakes,
  Hypocenter Location, and Shallow Fault Slip}.
\newblock \bibinfo{journal}{\emph{Front. in Earth Sci.}}  \bibinfo{volume}{9}
  (\bibinfo{year}{2021}), \bibinfo{pages}{177}.
\newblock


\bibitem[\protect\citeauthoryear{Berger, George, LeVeque, and Mandli}{Berger
  et~al\mbox{.}}{2011}]%
        {berger2011geoclaw}
\bibfield{author}{\bibinfo{person}{Marsha~J. Berger}, \bibinfo{person}{David~L.
  George}, \bibinfo{person}{Randall~J. LeVeque}, {and} \bibinfo{person}{Kyle~T.
  Mandli}.} \bibinfo{year}{2011}\natexlab{}.
\newblock \showarticletitle{The GeoClaw software for depth-averaged flows with
  adaptive refinement}.
\newblock \bibinfo{journal}{\emph{Adv. Water. Res.}} \bibinfo{volume}{34},
  \bibinfo{number}{9} (\bibinfo{year}{2011}), \bibinfo{pages}{1195--1206}.
\newblock


\bibitem[\protect\citeauthoryear{Breuer, Heinecke, and Bader}{Breuer
  et~al\mbox{.}}{2016}]%
        {breuer_petascale_2016}
\bibfield{author}{\bibinfo{person}{Alexander Breuer},
  \bibinfo{person}{Alexander Heinecke}, {and} \bibinfo{person}{Michael Bader}.}
  \bibinfo{year}{2016}\natexlab{}.
\newblock \showarticletitle{Petascale Local Time Stepping for the {ADER}-{DG}
  {Finite} {Element} Method}. In \bibinfo{booktitle}{\emph{2016 {IEEE} Int.
  Parallel Distrib. Process. Symp. ({IPDPS})}}. \bibinfo{pages}{854--863}.
\newblock
\urldef\tempurl%
\url{https://doi.org/10.1109/IPDPS.2016.109}
\showDOI{\tempurl}


\bibitem[\protect\citeauthoryear{Chen, Fu, Wei, He, Zhang, Li, Wan, Zhang, Gan,
  Zhang, Zhang, Yang, and Chen}{Chen et~al\mbox{.}}{2018}]%
        {chen:SC18}
\bibfield{author}{\bibinfo{person}{Bingwei Chen}, \bibinfo{person}{Haohuan Fu},
  \bibinfo{person}{Yanwen Wei}, \bibinfo{person}{Conghui He},
  \bibinfo{person}{Wenqiang Zhang}, \bibinfo{person}{Yuxuan Li},
  \bibinfo{person}{Wubin Wan}, \bibinfo{person}{Wei Zhang},
  \bibinfo{person}{Lin Gan}, \bibinfo{person}{Wei Zhang},
  \bibinfo{person}{Zhenguo Zhang}, \bibinfo{person}{Guangwen Yang}, {and}
  \bibinfo{person}{Xiaofei Chen}.} \bibinfo{year}{2018}\natexlab{}.
\newblock \showarticletitle{Simulating the Wenchuan Earthquake with Accurate
  Surface Topography on Sunway TaihuLight}. In \bibinfo{booktitle}{\emph{Proc.
  of the Int. Conf. High Perform. Comput., Networking, Storage, and Analysis}}
  (Dallas, Texas) \emph{(\bibinfo{series}{SC '18})}. Article
  \bibinfo{articleno}{40}.
\newblock
\urldef\tempurl%
\url{https://doi.org/10.1109/SC.2018.00043}
\showDOI{\tempurl}


\bibitem[\protect\citeauthoryear{Courant, Friedrichs, and Lewy}{Courant
  et~al\mbox{.}}{1928}]%
        {courant1928}
\bibfield{author}{\bibinfo{person}{Richard Courant}, \bibinfo{person}{Kurt
  Friedrichs}, {and} \bibinfo{person}{Hans Lewy}.}
  \bibinfo{year}{1928}\natexlab{}.
\newblock \showarticletitle{{\"U}ber die partiellen Differenzengleichungen der
  mathematischen Physik}.
\newblock \bibinfo{journal}{\emph{Math. Ann.}} \bibinfo{volume}{100},
  \bibinfo{number}{1} (\bibinfo{year}{1928}), \bibinfo{pages}{32--74}.
\newblock


\bibitem[\protect\citeauthoryear{DEMNAS}{DEMNAS}{2018}]%
        {DEMNAS2018}
DEMNAS \bibinfo{year}{2018}\natexlab{}.
\newblock \bibinfo{title}{{DEMNAS} -- Seamless Digital Elevation Model {(DEM)}
  dan Batimetri Nasional}.
\newblock \bibinfo{howpublished}{Badan Informasi Geospasial}.
\newblock
\urldef\tempurl%
\url{http://tides.big.go.id/DEMNAS}
\showURL{%
\tempurl}


\bibitem[\protect\citeauthoryear{Dumbser, Balsara, Toro, and Munz}{Dumbser
  et~al\mbox{.}}{2008}]%
        {dumbser_unified_2008}
\bibfield{author}{\bibinfo{person}{Michael Dumbser},
  \bibinfo{person}{Dinshaw~S. Balsara}, \bibinfo{person}{Eleuterio~F. Toro},
  {and} \bibinfo{person}{Claus-Dieter Munz}.} \bibinfo{year}{2008}\natexlab{}.
\newblock \showarticletitle{A unified framework for the construction of
  one-step finite volume and discontinuous {Galerkin} schemes on unstructured
  meshes}.
\newblock \bibinfo{journal}{\emph{J. Comput. Phys.}} \bibinfo{volume}{227},
  \bibinfo{number}{18} (\bibinfo{date}{Sept.} \bibinfo{year}{2008}),
  \bibinfo{pages}{8209--8253}.
\newblock
\urldef\tempurl%
\url{https://doi.org/10.1016/j.jcp.2008.05.025}
\showDOI{\tempurl}


\bibitem[\protect\citeauthoryear{Dumbser and Käser}{Dumbser and
  Käser}{2006}]%
        {dumbser_arbitrary_2006}
\bibfield{author}{\bibinfo{person}{Michael Dumbser} {and}
  \bibinfo{person}{Martin Käser}.} \bibinfo{year}{2006}\natexlab{}.
\newblock \showarticletitle{An arbitrary high-order discontinuous {Galerkin}
  method for elastic waves on unstructured meshes -- {II}. {The}
  three-dimensional isotropic case}.
\newblock \bibinfo{journal}{\emph{Geophys. J. Int.}} \bibinfo{volume}{167},
  \bibinfo{number}{1} (\bibinfo{date}{Oct.} \bibinfo{year}{2006}),
  \bibinfo{pages}{319--336}.
\newblock
\showISSN{0956540X, 1365246X}
\urldef\tempurl%
\url{https://doi.org/10.1111/j.1365-246X.2006.03120.x}
\showDOI{\tempurl}


\bibitem[\protect\citeauthoryear{Dumbser, Käser, and Toro}{Dumbser
  et~al\mbox{.}}{2007}]%
        {dumbser_arbitrary_2007}
\bibfield{author}{\bibinfo{person}{Michael Dumbser}, \bibinfo{person}{Martin
  Käser}, {and} \bibinfo{person}{Eleuterio~F. Toro}.}
  \bibinfo{year}{2007}\natexlab{}.
\newblock \showarticletitle{An arbitrary high-order {Discontinuous} {Galerkin}
  method for elastic waves on unstructured meshes -- {V}. {Local} time stepping
  and p-adaptivity}.
\newblock \bibinfo{journal}{\emph{Geophys. J. Int.}} \bibinfo{volume}{171},
  \bibinfo{number}{2} (\bibinfo{date}{Nov.} \bibinfo{year}{2007}),
  \bibinfo{pages}{695--717}.
\newblock
\showISSN{0956540X, 1365246X}
\urldef\tempurl%
\url{https://doi.org/10.1111/j.1365-246X.2007.03427.x}
\showDOI{\tempurl}


\bibitem[\protect\citeauthoryear{Dumbser, Schwartzkopff, and Munz}{Dumbser
  et~al\mbox{.}}{2006}]%
        {dumbserArbitraryHighOrder2006}
\bibfield{author}{\bibinfo{person}{M. Dumbser}, \bibinfo{person}{T.
  Schwartzkopff}, {and} \bibinfo{person}{C.-D. Munz}.}
  \bibinfo{year}{2006}\natexlab{}.
\newblock \showarticletitle{Arbitrary High Order Finite Volume Schemes for
  Linear Wave Propagation}. In \bibinfo{booktitle}{\emph{Comput. Sci. and High
  Perform. Comput. II}} \emph{(\bibinfo{series}{Notes on {{Numer. Fluid Mech.}}
  and {{Multidisci. Design}}})}. \bibinfo{publisher}{{Springer}},
  \bibinfo{address}{{Berlin, Heidelberg}}, \bibinfo{pages}{129--144}.
\newblock
\showISBNx{978-3-540-31768-5}
\urldef\tempurl%
\url{https://doi.org/10.1007/3-540-31768-6_11}
\showDOI{\tempurl}


\bibitem[\protect\citeauthoryear{Elbanna, Abdelmeguid, Ma, Amlani, Bhat,
  Synolakis, and Rosakis}{Elbanna et~al\mbox{.}}{2021}]%
        {Elbanna2021}
\bibfield{author}{\bibinfo{person}{Ahmed Elbanna}, \bibinfo{person}{Mohamed
  Abdelmeguid}, \bibinfo{person}{Xiao Ma}, \bibinfo{person}{Faisal Amlani},
  \bibinfo{person}{Harsha~S. Bhat}, \bibinfo{person}{Costas Synolakis}, {and}
  \bibinfo{person}{Ares~J. Rosakis}.} \bibinfo{year}{2021}\natexlab{}.
\newblock \showarticletitle{Anatomy of strike-slip fault tsunami genesis}.
\newblock \bibinfo{journal}{\emph{Proc. of the Natl. Acad. Sci.}}
  \bibinfo{volume}{118}, \bibinfo{number}{19} (\bibinfo{year}{2021}).
\newblock
\urldef\tempurl%
\url{https://doi.org/10.1073/pnas.2025632118}
\showDOI{\tempurl}


\bibitem[\protect\citeauthoryear{Fu, He, Chen, Yin, Zhang, Zhang, Zhang, Xue,
  Liu, Yin, Yang, and Chen}{Fu et~al\mbox{.}}{2017}]%
        {Fu:SC17}
\bibfield{author}{\bibinfo{person}{Haohuan Fu}, \bibinfo{person}{Conghui He},
  \bibinfo{person}{Bingwei Chen}, \bibinfo{person}{Zekun Yin},
  \bibinfo{person}{Zhenguo Zhang}, \bibinfo{person}{Wenqiang Zhang},
  \bibinfo{person}{Tingjian Zhang}, \bibinfo{person}{Wei Xue},
  \bibinfo{person}{Weiguo Liu}, \bibinfo{person}{Wanwang Yin},
  \bibinfo{person}{Guangwen Yang}, {and} \bibinfo{person}{Xiaofei Chen}.}
  \bibinfo{year}{2017}\natexlab{}.
\newblock \showarticletitle{18.9-pflops nonlinear earthquake simulation on
  {Sunway TaihuLight}: enabling depiction of 18-Hz and 8-meter scenarios}. In
  \bibinfo{booktitle}{\emph{Proc. Int. Conf. High Perform. Comput., Networking,
  Storage and Analysis}} (Denver, Colorado) \emph{(\bibinfo{series}{SC '17})}.
  \bibinfo{publisher}{Association for Computing Machinery},
  \bibinfo{address}{New York, NY, USA}, Article \bibinfo{articleno}{2},
  \bibinfo{numpages}{12}~pages.
\newblock
\showISBNx{9781450351140}
\urldef\tempurl%
\url{https://doi.org/10.1145/3126908.3126910}
\showDOI{\tempurl}


\bibitem[\protect\citeauthoryear{Harris, Barall, Aagaard, Ma, Roten, Olsen,
  Duan, Liu, Luo, Bai, Ampuero, Kaneko, Gabriel, Duru, Ulrich, Wollherr, Shi,
  Dunham, Bydlon, Zhang, Chen, Somala, Pelties, Tago, Cruz‐Atienza, Kozdon,
  Daub, Aslam, Kase, Withers, and Dalguer}{Harris et~al\mbox{.}}{2018}]%
        {harris_suite_2018}
\bibfield{author}{\bibinfo{person}{Ruth~A. Harris}, \bibinfo{person}{Michael
  Barall}, \bibinfo{person}{Brad Aagaard}, \bibinfo{person}{Shuo Ma},
  \bibinfo{person}{Daniel Roten}, \bibinfo{person}{Kim Olsen},
  \bibinfo{person}{Benchun Duan}, \bibinfo{person}{Dunyu Liu},
  \bibinfo{person}{Bin Luo}, \bibinfo{person}{Kangchen Bai},
  \bibinfo{person}{Jean‐Paul Ampuero}, \bibinfo{person}{Yoshihiro Kaneko},
  \bibinfo{person}{Alice‐Agnes Gabriel}, \bibinfo{person}{Kenneth Duru},
  \bibinfo{person}{Thomas Ulrich}, \bibinfo{person}{Stephanie Wollherr},
  \bibinfo{person}{Zheqiang Shi}, \bibinfo{person}{Eric Dunham},
  \bibinfo{person}{Sam Bydlon}, \bibinfo{person}{Zhenguo Zhang},
  \bibinfo{person}{Xiaofei Chen}, \bibinfo{person}{Surendra~Nadh Somala},
  \bibinfo{person}{Christian Pelties}, \bibinfo{person}{Josué Tago},
  \bibinfo{person}{Victor~Manuel Cruz‐Atienza}, \bibinfo{person}{Jeremy
  Kozdon}, \bibinfo{person}{Eric Daub}, \bibinfo{person}{Khurram Aslam},
  \bibinfo{person}{Yuko Kase}, \bibinfo{person}{Kyle Withers}, {and}
  \bibinfo{person}{Luis Dalguer}.} \bibinfo{year}{2018}\natexlab{}.
\newblock \showarticletitle{A {Suite} of {Exercises} for {Verifying} {Dynamic}
  {Earthquake} {Rupture} {Codes}}.
\newblock \bibinfo{journal}{\emph{Seismol. Res. Lett.}} \bibinfo{volume}{89},
  \bibinfo{number}{3} (\bibinfo{date}{May} \bibinfo{year}{2018}),
  \bibinfo{pages}{1146--1162}.
\newblock
\showISSN{0895-0695, 1938-2057}
\urldef\tempurl%
\url{https://doi.org/10.1785/0220170222}
\showDOI{\tempurl}


\bibitem[\protect\citeauthoryear{Heinecke, Breuer, Rettenberger, Bader,
  Gabriel, Pelties, Bode, Barth, Liao, Vaidyanathan, Smelyanskiy, and
  Dubey}{Heinecke et~al\mbox{.}}{2014}]%
        {heinecke_petascale_2014}
\bibfield{author}{\bibinfo{person}{Alexander Heinecke},
  \bibinfo{person}{Alexander Breuer}, \bibinfo{person}{Sebastian Rettenberger},
  \bibinfo{person}{Michael Bader}, \bibinfo{person}{Alice-Agnes Gabriel},
  \bibinfo{person}{Christian Pelties}, \bibinfo{person}{Arndt Bode},
  \bibinfo{person}{William Barth}, \bibinfo{person}{Xiang-Ke Liao},
  \bibinfo{person}{Karthikeyan Vaidyanathan}, \bibinfo{person}{Mikhail
  Smelyanskiy}, {and} \bibinfo{person}{Pradeep Dubey}.}
  \bibinfo{year}{2014}\natexlab{}.
\newblock \showarticletitle{Petascale High Order Dynamic Rupture Earthquake
  Simulations on Heterogeneous Supercomputers}. In
  \bibinfo{booktitle}{\emph{{SC} '14: Proc. Int. Conf. High Perform. Comp.,
  Networking, Storage and Analysis}}. \bibinfo{pages}{3--14}.
\newblock
\urldef\tempurl%
\url{https://doi.org/10.1109/SC.2014.6}
\showDOI{\tempurl}
\newblock
\shownote{ISSN: 2167-4337.}


\bibitem[\protect\citeauthoryear{Heinecke, Henry, Hutchinson, and
  Pabst}{Heinecke et~al\mbox{.}}{2016}]%
        {heinecke_libxsmm_2016}
\bibfield{author}{\bibinfo{person}{Alexander Heinecke}, \bibinfo{person}{Greg
  Henry}, \bibinfo{person}{Maxwell Hutchinson}, {and} \bibinfo{person}{Hans
  Pabst}.} \bibinfo{year}{2016}\natexlab{}.
\newblock \showarticletitle{LIBXSMM: accelerating small matrix multiplications
  by runtime code generation}. In \bibinfo{booktitle}{\emph{SC'16: Proc. Int.
  Conf. High Perform. Comput., Networking, Storage and Analysis}}. IEEE,
  \bibinfo{pages}{981--991}.
\newblock


\bibitem[\protect\citeauthoryear{Hoefler and Lumsdaine}{Hoefler and
  Lumsdaine}{2008}]%
        {hoefler2008}
\bibfield{author}{\bibinfo{person}{Torsten Hoefler} {and}
  \bibinfo{person}{Andrew Lumsdaine}.} \bibinfo{year}{2008}\natexlab{}.
\newblock \showarticletitle{Message progression in parallel computing -- to
  thread or not to thread?}. In \bibinfo{booktitle}{\emph{2008 IEEE Int. Conf.
  Cluster Comp.}} \bibinfo{pages}{213--222}.
\newblock
\urldef\tempurl%
\url{https://doi.org/10.1109/CLUSTR.2008.4663774}
\showDOI{\tempurl}


\bibitem[\protect\citeauthoryear{Ichimura, Fujita, Tanaka, Hori, Lalith,
  Shizawa, and Kobayashi}{Ichimura et~al\mbox{.}}{2014}]%
        {ichimura:SC14}
\bibfield{author}{\bibinfo{person}{Tsuyoshi Ichimura}, \bibinfo{person}{Kohei
  Fujita}, \bibinfo{person}{Seizo Tanaka}, \bibinfo{person}{Muneo Hori},
  \bibinfo{person}{Maddegedara Lalith}, \bibinfo{person}{Yoshihisa Shizawa},
  {and} \bibinfo{person}{Hiroshi Kobayashi}.} \bibinfo{year}{2014}\natexlab{}.
\newblock \showarticletitle{Physics-Based Urban Earthquake Simulation Enhanced
  by 10.7 BlnDOF $\times$ 30 K Time-Step Unstructured FE Non-Linear Seismic
  Wave Simulation}. In \bibinfo{booktitle}{\emph{Proc. Int. Conf. High Perform.
  Comput., Networking, Storage and Analysis}} \emph{(\bibinfo{series}{SC
  '14})}. \bibinfo{pages}{15–26}.
\newblock
\showISBNx{9781479955008}
\urldef\tempurl%
\url{https://doi.org/10.1109/SC.2014.7}
\showDOI{\tempurl}


\bibitem[\protect\citeauthoryear{Ichimura, Fujita, Yamaguchi, Naruse, Wells,
  Schulthess, Straatsma, Zimmer, Martinasso, Nakajima, Hori, and
  Maddegedara}{Ichimura et~al\mbox{.}}{2018}]%
        {ichimura:SC18}
\bibfield{author}{\bibinfo{person}{Tsuyoshi Ichimura}, \bibinfo{person}{Kohei
  Fujita}, \bibinfo{person}{Takuma Yamaguchi}, \bibinfo{person}{Akira Naruse},
  \bibinfo{person}{Jack~C. Wells}, \bibinfo{person}{Thomas~C. Schulthess},
  \bibinfo{person}{Tjerk~P. Straatsma}, \bibinfo{person}{Christopher~J.
  Zimmer}, \bibinfo{person}{Maxime Martinasso}, \bibinfo{person}{Kengo
  Nakajima}, \bibinfo{person}{Muneo Hori}, {and} \bibinfo{person}{Lalith
  Maddegedara}.} \bibinfo{year}{2018}\natexlab{}.
\newblock \showarticletitle{A fast scalable implicit solver for nonlinear
  time-evolution earthquake city problem on low-ordered unstructured finite
  elements with artificial intelligence and transprecision computing}. In
  \bibinfo{booktitle}{\emph{SC18: Int. Conf. High Perform. Comput., Networking,
  Storage and Analysis}}. IEEE, \bibinfo{pages}{627--637}.
\newblock


\bibitem[\protect\citeauthoryear{Kajiura}{Kajiura}{1963}]%
        {kajiura1963}
\bibfield{author}{\bibinfo{person}{Kinjiro Kajiura}.}
  \bibinfo{year}{1963}\natexlab{}.
\newblock \showarticletitle{The leading wave of a tsunami}.
\newblock \bibinfo{journal}{\emph{Bull. Earthquake Res. Inst., University of
  Tokyo}} \bibinfo{volume}{41}, \bibinfo{number}{3} (\bibinfo{year}{1963}),
  \bibinfo{pages}{535--571}.
\newblock


\bibitem[\protect\citeauthoryear{K{\"{a}}ser and Dumbser}{K{\"{a}}ser and
  Dumbser}{2006}]%
        {Kaser2006}
\bibfield{author}{\bibinfo{person}{Martin K{\"{a}}ser} {and}
  \bibinfo{person}{Michael Dumbser}.} \bibinfo{year}{2006}\natexlab{}.
\newblock \showarticletitle{{An arbitrary high-order discontinuous Galerkin
  method for elastic waves on unstructured meshes -- I. The two-dimensional
  isotropic case with external source terms}}.
\newblock \bibinfo{journal}{\emph{Geophysical J.\ Int.}} \bibinfo{volume}{166},
  \bibinfo{number}{2} (\bibinfo{date}{Aug} \bibinfo{year}{2006}),
  \bibinfo{pages}{855--877}.
\newblock
\urldef\tempurl%
\url{https://doi.org/10.1111/j.1365-246X.2006.03051.x}
\showDOI{\tempurl}


\bibitem[\protect\citeauthoryear{Klinkenberg, Samfass, Bader, Terboven, and
  M\"{u}ller}{Klinkenberg et~al\mbox{.}}{2020}]%
        {klinkenberg2019}
\bibfield{author}{\bibinfo{person}{Jannis Klinkenberg},
  \bibinfo{person}{Philipp Samfass}, \bibinfo{person}{Michael Bader},
  \bibinfo{person}{Christian Terboven}, {and} \bibinfo{person}{Matthias~S.
  M\"{u}ller}.} \bibinfo{year}{2020}\natexlab{}.
\newblock \showarticletitle{Reactive task migration for hybrid {MPI} + {OpenMP}
  applications}. In \bibinfo{booktitle}{\emph{Parallel Proc. and Appl. Math.}}
  \bibinfo{publisher}{Springer International Publishing},
  \bibinfo{pages}{59--71}.
\newblock
\urldef\tempurl%
\url{https://doi.org/10.1007/978-3-030-43222-5_6}
\showDOI{\tempurl}


\bibitem[\protect\citeauthoryear{Komatitsch}{Komatitsch}{2011}]%
        {komatitsch2011}
\bibfield{author}{\bibinfo{person}{Dimitri Komatitsch}.}
  \bibinfo{year}{2011}\natexlab{}.
\newblock \showarticletitle{Fluid--solid coupling on a cluster of GPU graphics
  cards for seismic wave propagation}.
\newblock \bibinfo{journal}{\emph{Comptes Rendus M{\'e}canique}}
  \bibinfo{volume}{339}, \bibinfo{number}{2-3} (\bibinfo{year}{2011}),
  \bibinfo{pages}{125--135}.
\newblock


\bibitem[\protect\citeauthoryear{Kozdon and Dunham}{Kozdon and Dunham}{2014}]%
        {KozdonDunham2014}
\bibfield{author}{\bibinfo{person}{Jeremy~E. Kozdon} {and}
  \bibinfo{person}{Eric~M. Dunham}.} \bibinfo{year}{2014}\natexlab{}.
\newblock \showarticletitle{Constraining shallow slip and tsunami excitation in
  megathrust ruptures using seismic and ocean acoustic waves recorded on
  ocean-bottom sensor networks}.
\newblock \bibinfo{journal}{\emph{Earth Planet. Sci. Lett.}}
  \bibinfo{volume}{396} (\bibinfo{year}{2014}), \bibinfo{pages}{56--65}.
\newblock
\showISSN{0012-821X}
\urldef\tempurl%
\url{https://doi.org/10.1016/j.epsl.2014.04.001}
\showDOI{\tempurl}


\bibitem[\protect\citeauthoryear{Krenz, Uphoff, Abrahams, Gabriel, Dunham, and
  Bader}{Krenz et~al\mbox{.}}{2019}]%
        {krenz_elastic-acoustic_2019}
\bibfield{author}{\bibinfo{person}{Lukas Krenz}, \bibinfo{person}{Carsten
  Uphoff}, \bibinfo{person}{Lauren~S Abrahams}, \bibinfo{person}{Alice-Agnes
  Gabriel}, \bibinfo{person}{Eric~M Dunham}, {and} \bibinfo{person}{Michael
  Bader}.} \bibinfo{year}{2019}\natexlab{}.
\newblock \showarticletitle{Elastic-acoustic coupling for 3D tsunamigenic
  earthquake simulations with ADER-DG on unstructured tetrahedral meshes}. In
  \bibinfo{booktitle}{\emph{AGU Fall Meet. Abstr.}},
  Vol.~\bibinfo{volume}{2019}. \bibinfo{pages}{T52C--09}.
\newblock


\bibitem[\protect\citeauthoryear{Käser, Dumbser, De~La~Puente, and
  Igel}{Käser et~al\mbox{.}}{2007}]%
        {puente_arbitrary_2007}
\bibfield{author}{\bibinfo{person}{Martin Käser}, \bibinfo{person}{Michael
  Dumbser}, \bibinfo{person}{Josep De~La~Puente}, {and} \bibinfo{person}{Heiner
  Igel}.} \bibinfo{year}{2007}\natexlab{}.
\newblock \showarticletitle{{An arbitrary high-order Discontinuous Galerkin
  method for elastic waves on unstructured meshes — III. Viscoelastic
  attenuation}}.
\newblock \bibinfo{journal}{\emph{Geophys. J. Int.}} \bibinfo{volume}{168},
  \bibinfo{number}{1} (\bibinfo{date}{01} \bibinfo{year}{2007}),
  \bibinfo{pages}{224--242}.
\newblock
\showISSN{0956-540X}
\urldef\tempurl%
\url{https://doi.org/10.1111/j.1365-246X.2006.03193.x}
\showDOI{\tempurl}


\bibitem[\protect\citeauthoryear{Käser, Hermann, and Puente}{Käser
  et~al\mbox{.}}{2008}]%
        {kaser_quantitative_2008}
\bibfield{author}{\bibinfo{person}{Martin Käser}, \bibinfo{person}{Verena
  Hermann}, {and} \bibinfo{person}{Josep de~la Puente}.}
  \bibinfo{year}{2008}\natexlab{}.
\newblock \showarticletitle{Quantitative accuracy analysis of the discontinuous
  {Galerkin} method for seismic wave propagation}.
\newblock \bibinfo{journal}{\emph{Geophys. J. Int.}} \bibinfo{volume}{173},
  \bibinfo{number}{3} (\bibinfo{date}{June} \bibinfo{year}{2008}),
  \bibinfo{pages}{990--999}.
\newblock
\showISSN{0956540X, 1365246X}
\urldef\tempurl%
\url{https://doi.org/10.1111/j.1365-246X.2008.03781.x}
\showDOI{\tempurl}


\bibitem[\protect\citeauthoryear{LeVeque}{LeVeque}{2002}]%
        {leveque_finite_2002}
\bibfield{author}{\bibinfo{person}{Randall~J. LeVeque}.}
  \bibinfo{year}{2002}\natexlab{}.
\newblock \bibinfo{booktitle}{\emph{Finite {Volume} {Methods} for {Hyperbolic}
  {Problems}}}.
\newblock \bibinfo{publisher}{Cambridge University Press}.
\newblock
\urldef\tempurl%
\url{https://doi.org/10.1017/cbo9780511791253}
\showURL{%
\tempurl}


\bibitem[\protect\citeauthoryear{Liang and Dunham}{Liang and Dunham}{2020}]%
        {liang2020lava}
\bibfield{author}{\bibinfo{person}{Chao Liang} {and} \bibinfo{person}{Eric~M
  Dunham}.} \bibinfo{year}{2020}\natexlab{}.
\newblock \showarticletitle{Lava lake sloshing modes during the 2018
  K{\=\i}lauea Volcano eruption probe magma reservoir storativity}.
\newblock \bibinfo{journal}{\emph{Earth Planet. Sci. Lett.}}
  \bibinfo{volume}{535} (\bibinfo{year}{2020}), \bibinfo{pages}{116110}.
\newblock


\bibitem[\protect\citeauthoryear{Lotto and Dunham}{Lotto and Dunham}{2015}]%
        {lotto_high-order_2015}
\bibfield{author}{\bibinfo{person}{Gabriel~C. Lotto} {and}
  \bibinfo{person}{Eric~M. Dunham}.} \bibinfo{year}{2015}\natexlab{}.
\newblock \showarticletitle{High-order finite difference modeling of tsunami
  generation in a compressible ocean from offshore earthquakes}.
\newblock \bibinfo{journal}{\emph{Comput. Geosci.}} \bibinfo{volume}{19},
  \bibinfo{number}{2} (\bibinfo{year}{2015}), \bibinfo{pages}{327--340}.
\newblock
\showISSN{1420-0597, 1573-1499}
\urldef\tempurl%
\url{https://doi.org/10.1007/s10596-015-9472-0}
\showDOI{\tempurl}


\bibitem[\protect\citeauthoryear{Lotto, Jeppson, and Dunham}{Lotto
  et~al\mbox{.}}{2019}]%
        {lotto2018fully}
\bibfield{author}{\bibinfo{person}{Gabriel~C. Lotto},
  \bibinfo{person}{Tamara~N. Jeppson}, {and} \bibinfo{person}{Eric~M. Dunham}.}
  \bibinfo{year}{2019}\natexlab{}.
\newblock \showarticletitle{Fully coupled simulations of megathrust earthquakes
  and tsunamis in the Japan Trench, Nankai Trough, and Cascadia Subduction
  Zone}.
\newblock \bibinfo{journal}{\emph{Pure Appl. Geophys.}} \bibinfo{volume}{176},
  \bibinfo{number}{9} (\bibinfo{year}{2019}), \bibinfo{pages}{4009--4041}.
\newblock


\bibitem[\protect\citeauthoryear{Lotto, Nava, and Dunham}{Lotto
  et~al\mbox{.}}{2017}]%
        {lotto_should_2017}
\bibfield{author}{\bibinfo{person}{Gabriel~C. Lotto}, \bibinfo{person}{Gabriel
  Nava}, {and} \bibinfo{person}{Eric~M. Dunham}.}
  \bibinfo{year}{2017}\natexlab{}.
\newblock \showarticletitle{Should tsunami simulations include a nonzero
  initial horizontal velocity?}
\newblock \bibinfo{journal}{\emph{Earth Planets Space}} \bibinfo{volume}{69},
  \bibinfo{number}{1} (\bibinfo{date}{Dec.} \bibinfo{year}{2017}),
  \bibinfo{pages}{117}.
\newblock
\showISSN{1880-5981}
\urldef\tempurl%
\url{https://doi.org/10.1186/s40623-017-0701-8}
\showDOI{\tempurl}


\bibitem[\protect\citeauthoryear{Madden, Bader, Behrens, van Dinther, Gabriel,
  Rannabauer, Ulrich, Uphoff, Vater, and van Zelst}{Madden
  et~al\mbox{.}}{2021}]%
        {madden_linked_2020}
\bibfield{author}{\bibinfo{person}{E.~H. Madden}, \bibinfo{person}{M. Bader},
  \bibinfo{person}{J. Behrens}, \bibinfo{person}{Y. van Dinther},
  \bibinfo{person}{A.-A. Gabriel}, \bibinfo{person}{L. Rannabauer},
  \bibinfo{person}{T. Ulrich}, \bibinfo{person}{C. Uphoff}, \bibinfo{person}{S.
  Vater}, {and} \bibinfo{person}{I. van Zelst}.}
  \bibinfo{year}{2021}\natexlab{}.
\newblock \showarticletitle{Linked 3-{D} modelling of megathrust
  earthquake-tsunami events: from subduction to tsunami run up}.
\newblock \bibinfo{journal}{\emph{Geophys. J. Int.}} \bibinfo{volume}{224},
  \bibinfo{number}{1} (\bibinfo{year}{2021}), \bibinfo{pages}{487--516}.
\newblock
\showISSN{0956-540X}
\urldef\tempurl%
\url{https://doi.org/10.1093/gji/ggaa484}
\showDOI{\tempurl}


\bibitem[\protect\citeauthoryear{Maeda, Furumura, Noguchi, Takemura, Sakai,
  Shinohara, Iwai, and Lee}{Maeda et~al\mbox{.}}{2013}]%
        {Maeda2013}
\bibfield{author}{\bibinfo{person}{Takuto Maeda}, \bibinfo{person}{Takashi
  Furumura}, \bibinfo{person}{Shinako Noguchi}, \bibinfo{person}{Shunsuke
  Takemura}, \bibinfo{person}{Shin'ichi Sakai}, \bibinfo{person}{Masanao
  Shinohara}, \bibinfo{person}{Kazuhisa Iwai}, {and}
  \bibinfo{person}{Shiann~Jong Lee}.} \bibinfo{year}{2013}\natexlab{}.
\newblock \showarticletitle{{Seismic- and Tsunami-wave propagation of the 2011
  Off the Pacific Coast of Tohoku earthquake as inferred from the
  Tsunami-coupled finite-difference simulation}}.
\newblock \bibinfo{journal}{\emph{Bull. Seismol. Soc. Am.}}
  \bibinfo{volume}{103}, \bibinfo{number}{2 B} (\bibinfo{date}{may}
  \bibinfo{year}{2013}), \bibinfo{pages}{1456--1472}.
\newblock
\showISSN{00371106}
\urldef\tempurl%
\url{https://doi.org/10.1785/0120120118}
\showDOI{\tempurl}


\bibitem[\protect\citeauthoryear{Mai}{Mai}{2019}]%
        {maiSupershearTsunamiDisaster2019}
\bibfield{author}{\bibinfo{person}{P.~Martin Mai}.}
  \bibinfo{year}{2019}\natexlab{}.
\newblock \showarticletitle{Supershear tsunami disaster}.
\newblock \bibinfo{journal}{\emph{Nat. Geosci.}} \bibinfo{volume}{12},
  \bibinfo{number}{3} (\bibinfo{date}{March} \bibinfo{year}{2019}),
  \bibinfo{pages}{150--151}.
\newblock
\showISSN{1752-0908}
\urldef\tempurl%
\url{https://doi.org/10.1038/s41561-019-0308-8}
\showDOI{\tempurl}


\bibitem[\protect\citeauthoryear{McCalpin}{McCalpin}{2018}]%
        {McCalpin:sc18}
\bibfield{author}{\bibinfo{person}{John~D. McCalpin}.}
  \bibinfo{year}{2018}\natexlab{}.
\newblock \showarticletitle{{HPL} and {DGEMM} performance variability on the
  {Xeon Platinum 8160} Processor}. In \bibinfo{booktitle}{\emph{Proc. of the
  Int. Conf. High Perform. Comput., Networking, Storage, and Analysis}}
  \emph{(\bibinfo{series}{SC '18})}.
\newblock
\urldef\tempurl%
\url{https://doi.org/10.1109/SC.2018.00021}
\showDOI{\tempurl}


\bibitem[\protect\citeauthoryear{Milner, Shaw, Goulet, Richards‐Dinger,
  Callaghan, Jordan, Dieterich, and Field}{Milner et~al\mbox{.}}{2021}]%
        {Milner2021}
\bibfield{author}{\bibinfo{person}{Kevin~R. Milner}, \bibinfo{person}{Bruce~E.
  Shaw}, \bibinfo{person}{Christine~A. Goulet}, \bibinfo{person}{Keith~B.
  Richards‐Dinger}, \bibinfo{person}{Scott Callaghan},
  \bibinfo{person}{Thomas~H. Jordan}, \bibinfo{person}{James~H. Dieterich},
  {and} \bibinfo{person}{Edward~H. Field}.} \bibinfo{year}{2021}\natexlab{}.
\newblock \showarticletitle{{Toward Physics‐Based Nonergodic PSHA: A
  Prototype Fully Deterministic Seismic Hazard Model for Southern California}}.
\newblock \bibinfo{journal}{\emph{Bull. Seismol. Soc. Am.}}
  \bibinfo{volume}{111}, \bibinfo{number}{2} (\bibinfo{date}{01}
  \bibinfo{year}{2021}), \bibinfo{pages}{898--915}.
\newblock


\bibitem[\protect\citeauthoryear{Okada}{Okada}{1985}]%
        {okada_surface_1985}
\bibfield{author}{\bibinfo{person}{Yoshimitsu Okada}.}
  \bibinfo{year}{1985}\natexlab{}.
\newblock \showarticletitle{Surface deformation due to shear and tensile faults
  in a half-space}.
\newblock \bibinfo{journal}{\emph{Bull. Seismol. Soc. Am.}}
  \bibinfo{volume}{75}, \bibinfo{number}{4} (\bibinfo{date}{Aug.}
  \bibinfo{year}{1985}), \bibinfo{pages}{1135--1154}.
\newblock
\showISSN{0037-1106}
\urldef\tempurl%
\url{https://doi.org/10.1785/BSSA0750041135}
\showURL{%
\tempurl}


\bibitem[\protect\citeauthoryear{Pelties, Gabriel, and Ampuero}{Pelties
  et~al\mbox{.}}{2014}]%
        {pelties_gmd_2014}
\bibfield{author}{\bibinfo{person}{C. Pelties}, \bibinfo{person}{A.-A.
  Gabriel}, {and} \bibinfo{person}{J.-P. Ampuero}.}
  \bibinfo{year}{2014}\natexlab{}.
\newblock \showarticletitle{Verification of an {ADER-DG} method for complex
  dynamic rupture problems}.
\newblock \bibinfo{journal}{\emph{Geosci. Model Dev.}} \bibinfo{volume}{7},
  \bibinfo{number}{3} (\bibinfo{year}{2014}), \bibinfo{pages}{847--866}.
\newblock
\urldef\tempurl%
\url{https://doi.org/10.5194/gmd-7-847-2014}
\showDOI{\tempurl}


\bibitem[\protect\citeauthoryear{Rettenberger and Bader}{Rettenberger and
  Bader}{2015}]%
        {rettenberger_optimizing_2015}
\bibfield{author}{\bibinfo{person}{Sebastian Rettenberger} {and}
  \bibinfo{person}{Michael Bader}.} \bibinfo{year}{2015}\natexlab{}.
\newblock \showarticletitle{Optimizing {I/O} for petascale seismic simulations
  on unstructured meshes}. In \bibinfo{booktitle}{\emph{2015 IEEE Int. Conf.
  Cluster Comp.}} IEEE, \bibinfo{pages}{314--317}.
\newblock
\urldef\tempurl%
\url{https://doi.org/10.1109/CLUSTER.2015.51}
\showDOI{\tempurl}


\bibitem[\protect\citeauthoryear{Rietmann, Grote, Peter, and Schenk}{Rietmann
  et~al\mbox{.}}{2017}]%
        {rietmann2017}
\bibfield{author}{\bibinfo{person}{Max Rietmann}, \bibinfo{person}{Marcus
  Grote}, \bibinfo{person}{Daniel Peter}, {and} \bibinfo{person}{Olaf Schenk}.}
  \bibinfo{year}{2017}\natexlab{}.
\newblock \showarticletitle{Newmark local time stepping on high-performance
  computing architectures}.
\newblock \bibinfo{journal}{\emph{J. Comput. Phys.}}  \bibinfo{volume}{334}
  (\bibinfo{year}{2017}), \bibinfo{pages}{308--326}.
\newblock


\bibitem[\protect\citeauthoryear{Rietmann, Peter, Schenk, U{\c{c}}ar, and
  Grote}{Rietmann et~al\mbox{.}}{2015}]%
        {rietmann2015}
\bibfield{author}{\bibinfo{person}{Max Rietmann}, \bibinfo{person}{Daniel
  Peter}, \bibinfo{person}{Olaf Schenk}, \bibinfo{person}{Bora U{\c{c}}ar},
  {and} \bibinfo{person}{Marcus Grote}.} \bibinfo{year}{2015}\natexlab{}.
\newblock \showarticletitle{Load-balanced local time stepping for large-scale
  wave propagation}. In \bibinfo{booktitle}{\emph{2015 IEEE Int. Parallel and
  Distrib. Process. Symp.}} IEEE, \bibinfo{pages}{925--935}.
\newblock


\bibitem[\protect\citeauthoryear{Rodgers, Pitarka, Pankajakshan, Sj{\"o}green,
  and Petersson}{Rodgers et~al\mbox{.}}{2020}]%
        {Rodgers2020}
\bibfield{author}{\bibinfo{person}{Arthur~J. Rodgers}, \bibinfo{person}{Arben
  Pitarka}, \bibinfo{person}{Ramesh Pankajakshan}, \bibinfo{person}{Bjorn
  Sj{\"o}green}, {and} \bibinfo{person}{N.~Anders Petersson}.}
  \bibinfo{year}{2020}\natexlab{}.
\newblock \showarticletitle{Regional-Scale 3D Ground-Motion Simulations of $M_w
  7$ Earthquakes on the Hayward Fault, Northern California Resolving
  Frequencies 0--10 Hz and Including Site-Response Corrections}.
\newblock \bibinfo{journal}{\emph{Bull. Seismol. Soc. Am.}}
  \bibinfo{volume}{110}, \bibinfo{number}{6} (\bibinfo{year}{2020}),
  \bibinfo{pages}{2862--2881}.
\newblock


\bibitem[\protect\citeauthoryear{Rodgers, Pitarka, Petersson, Sjögreen, and
  McCallen}{Rodgers et~al\mbox{.}}{2018}]%
        {Rodgers2018}
\bibfield{author}{\bibinfo{person}{Arthur~J. Rodgers}, \bibinfo{person}{Arben
  Pitarka}, \bibinfo{person}{N.~Anders Petersson}, \bibinfo{person}{Björn
  Sjögreen}, {and} \bibinfo{person}{David~B. McCallen}.}
  \bibinfo{year}{2018}\natexlab{}.
\newblock \showarticletitle{Broadband (0–4 Hz) Ground Motions for a Magnitude
  7.0 Hayward Fault Earthquake With Three-Dimensional Structure and
  Topography}.
\newblock \bibinfo{journal}{\emph{Geophys. Res. Lett.}} \bibinfo{volume}{45},
  \bibinfo{number}{2} (\bibinfo{year}{2018}), \bibinfo{pages}{739--747}.
\newblock
\urldef\tempurl%
\url{https://doi.org/10.1002/2017GL076505}
\showDOI{\tempurl}


\bibitem[\protect\citeauthoryear{Ronan~Madec}{Ronan~Madec}{2009}]%
        {madec2009}
\bibfield{author}{\bibinfo{person}{Julien~Diaz Ronan~Madec,
  Dimitri~Komatitsch}.} \bibinfo{year}{2009}\natexlab{}.
\newblock \showarticletitle{Energy-conserving local time stepping based on
  high-order finite elements for seismic wave propagation across a fluid-solid
  interface}.
\newblock \bibinfo{journal}{\emph{Comp. Model. in Eng. \& Sci.}}
  \bibinfo{volume}{49}, \bibinfo{number}{2} (\bibinfo{year}{2009}),
  \bibinfo{pages}{163--190}.
\newblock
\showISSN{1526-1506}
\urldef\tempurl%
\url{https://doi.org/10.3970/cmes.2009.049.163}
\showDOI{\tempurl}


\bibitem[\protect\citeauthoryear{Rosenkrantz, Bottero, Komatitsch, and
  Monteiller}{Rosenkrantz et~al\mbox{.}}{2019}]%
        {Rosenkrantz2019}
\bibfield{author}{\bibinfo{person}{Eric Rosenkrantz}, \bibinfo{person}{Alexis
  Bottero}, \bibinfo{person}{Dimitri Komatitsch}, {and} \bibinfo{person}{Vadim
  Monteiller}.} \bibinfo{year}{2019}\natexlab{}.
\newblock \showarticletitle{A flexible numerical approach for non-destructive
  ultrasonic testing based on a time-domain spectral-element method: Ultrasonic
  modeling of Lamb waves in immersed defective structures and of bulk waves in
  damaged anisotropic materials}.
\newblock \bibinfo{journal}{\emph{NDT \& E International}}
  \bibinfo{volume}{101} (\bibinfo{year}{2019}), \bibinfo{pages}{72--86}.
\newblock
\showISSN{0963-8695}
\urldef\tempurl%
\url{https://doi.org/10.1016/j.ndteint.2018.10.002}
\showDOI{\tempurl}


\bibitem[\protect\citeauthoryear{Roten, Cui, Olsen, Day, Withers, Savran, Wang,
  and Mu}{Roten et~al\mbox{.}}{2016}]%
        {Roten:SC16}
\bibfield{author}{\bibinfo{person}{Daniel Roten}, \bibinfo{person}{Yifeng Cui},
  \bibinfo{person}{Kim~B Olsen}, \bibinfo{person}{Steven~M Day},
  \bibinfo{person}{Kyle Withers}, \bibinfo{person}{William~H Savran},
  \bibinfo{person}{Peng Wang}, {and} \bibinfo{person}{Dawei Mu}.}
  \bibinfo{year}{2016}\natexlab{}.
\newblock \showarticletitle{High-frequency nonlinear earthquake simulations on
  petascale heterogeneous supercomputers}. In \bibinfo{booktitle}{\emph{SC'16:
  Proc. Int. Conf. High Perform. Comput., Networking, Storage and Analysis}}.
  IEEE, \bibinfo{pages}{957--968}.
\newblock


\bibitem[\protect\citeauthoryear{Saito, Baba, Inazu, Takemura, and
  Fukuyama}{Saito et~al\mbox{.}}{2019}]%
        {Saito2019}
\bibfield{author}{\bibinfo{person}{Tatsuhiko Saito}, \bibinfo{person}{Toshitaka
  Baba}, \bibinfo{person}{Daisuke Inazu}, \bibinfo{person}{Shunsuke Takemura},
  {and} \bibinfo{person}{Eiichi Fukuyama}.} \bibinfo{year}{2019}\natexlab{}.
\newblock \showarticletitle{Synthesizing sea surface height change including
  seismic waves and tsunami using a dynamic rupture scenario of anticipated
  {Nankai} trough earthquakes}.
\newblock \bibinfo{journal}{\emph{Tectonophys.}} (\bibinfo{year}{2019}),
  \bibinfo{pages}{228166}.
\newblock
\showISSN{0040-1951}
\urldef\tempurl%
\url{https://doi.org/10.1016/j.tecto.2019.228166}
\showDOI{\tempurl}


\bibitem[\protect\citeauthoryear{Schloegel, Karypis, and Kumar}{Schloegel
  et~al\mbox{.}}{2002}]%
        {schloegel2002parallel}
\bibfield{author}{\bibinfo{person}{Kirk Schloegel}, \bibinfo{person}{George
  Karypis}, {and} \bibinfo{person}{Vipin Kumar}.}
  \bibinfo{year}{2002}\natexlab{}.
\newblock \showarticletitle{Parallel static and dynamic multi-constraint graph
  partitioning}.
\newblock \bibinfo{journal}{\emph{Concurrency Comp.: Pract. and Exp.}}
  \bibinfo{volume}{14}, \bibinfo{number}{3} (\bibinfo{year}{2002}),
  \bibinfo{pages}{219--240}.
\newblock


\bibitem[\protect\citeauthoryear{Shi, Kirby, Harris, Geiman, and Grilli}{Shi
  et~al\mbox{.}}{2012}]%
        {Shi2012}
\bibfield{author}{\bibinfo{person}{Fengyan Shi}, \bibinfo{person}{James~T.
  Kirby}, \bibinfo{person}{Jeffrey~C. Harris}, \bibinfo{person}{Joseph~D.
  Geiman}, {and} \bibinfo{person}{Stephan~T. Grilli}.}
  \bibinfo{year}{2012}\natexlab{}.
\newblock \showarticletitle{{A high-order adaptive time-stepping TVD solver for
  Boussinesq modeling of breaking waves and coastal inundation}}.
\newblock \bibinfo{journal}{\emph{Ocean Modelling}}  \bibinfo{volume}{43-44}
  (\bibinfo{date}{Jan} \bibinfo{year}{2012}), \bibinfo{pages}{36--51}.
\newblock
\showISSN{14635003}
\urldef\tempurl%
\url{https://doi.org/10.1016/j.ocemod.2011.12.004}
\showDOI{\tempurl}


\bibitem[\protect\citeauthoryear{Sladen, Rivet, Ampuero, De~Barros, Hello,
  Calbris, and Lamare}{Sladen et~al\mbox{.}}{2019}]%
        {Sladen2019}
\bibfield{author}{\bibinfo{person}{Anthony Sladen}, \bibinfo{person}{Diane
  Rivet}, \bibinfo{person}{Jean-Paul Ampuero}, \bibinfo{person}{Louis
  De~Barros}, \bibinfo{person}{Yann Hello}, \bibinfo{person}{Ga{\"e}tan
  Calbris}, {and} \bibinfo{person}{Patrick Lamare}.}
  \bibinfo{year}{2019}\natexlab{}.
\newblock \showarticletitle{Distributed sensing of earthquakes and ocean-solid
  Earth interactions on seafloor telecom cables}.
\newblock \bibinfo{journal}{\emph{Nat. Commun.}} \bibinfo{volume}{10},
  \bibinfo{number}{1} (\bibinfo{year}{2019}), \bibinfo{pages}{1--8}.
\newblock


\bibitem[\protect\citeauthoryear{Sochacki, George, Ewing, and
  Smithson}{Sochacki et~al\mbox{.}}{1991}]%
        {sochacki_1991}
\bibfield{author}{\bibinfo{person}{J.~S. Sochacki}, \bibinfo{person}{J.~H.
  George}, \bibinfo{person}{R.~E. Ewing}, {and} \bibinfo{person}{S.~B.
  Smithson}.} \bibinfo{year}{1991}\natexlab{}.
\newblock \showarticletitle{Interface conditions for acoustic and elastic wave
  propagation}.
\newblock \bibinfo{journal}{\emph{GEOPHYSICS}} \bibinfo{volume}{56},
  \bibinfo{number}{2} (\bibinfo{year}{1991}), \bibinfo{pages}{168--181}.
\newblock
\urldef\tempurl%
\url{https://doi.org/10.1190/1.1443029}
\showDOI{\tempurl}


\bibitem[\protect\citeauthoryear{Song, Fu, Zlotnicki, Ji, Hjorleifsdottir,
  Shum, and Yi}{Song et~al\mbox{.}}{2008}]%
        {Song2008}
\bibfield{author}{\bibinfo{person}{Y.~Tony Song}, \bibinfo{person}{L.-L. Fu},
  \bibinfo{person}{Victor Zlotnicki}, \bibinfo{person}{Chen Ji},
  \bibinfo{person}{Vala Hjorleifsdottir}, \bibinfo{person}{C.K. Shum}, {and}
  \bibinfo{person}{Yuchan Yi}.} \bibinfo{year}{2008}\natexlab{}.
\newblock \showarticletitle{The role of horizontal impulses of the faulting
  continental slope in generating the 26 December 2004 tsunami}.
\newblock \bibinfo{journal}{\emph{Ocean Modell.}} \bibinfo{volume}{20},
  \bibinfo{number}{4} (\bibinfo{year}{2008}), \bibinfo{pages}{362--379}.
\newblock
\urldef\tempurl%
\url{https://doi.org/10.1016/j.ocemod.2007.10.007}
\showDOI{\tempurl}


\bibitem[\protect\citeauthoryear{Stephenson, Reitman, and Angster}{Stephenson
  et~al\mbox{.}}{2017}]%
        {Stephenson2017}
\bibfield{author}{\bibinfo{person}{W.J. Stephenson}, \bibinfo{person}{N.G.
  Reitman}, {and} \bibinfo{person}{S.J. Angster}.}
  \bibinfo{year}{2017}\natexlab{}.
\newblock \bibinfo{booktitle}{\emph{{U.S. Geological Survey Open-File Report
  2017--1152: P- and S-wave velocity models incorporating the Cascadia
  subduction zone for 3D earthquake ground motion simulations, version
  1.6---Update for Open-File Report 2007--1348}}}.
\newblock \bibinfo{type}{{T}echnical {R}eport}. \bibinfo{institution}{U.S.
  Geol. Surv.}
\newblock
\urldef\tempurl%
\url{https://doi.org/10.3133/ofr20171152}
\showDOI{\tempurl}


\bibitem[\protect\citeauthoryear{Titarev and Toro}{Titarev and Toro}{2002}]%
        {titarev2002ader}
\bibfield{author}{\bibinfo{person}{Vladimir~A. Titarev} {and}
  \bibinfo{person}{Eleuterio~F. Toro}.} \bibinfo{year}{2002}\natexlab{}.
\newblock \showarticletitle{ADER: Arbitrary high order Godunov approach}.
\newblock \bibinfo{journal}{\emph{J. Sci. Comput.}} \bibinfo{volume}{17},
  \bibinfo{number}{1} (\bibinfo{year}{2002}), \bibinfo{pages}{609--618}.
\newblock


\bibitem[\protect\citeauthoryear{Toro}{Toro}{2009}]%
        {toro_riemann_2009}
\bibfield{author}{\bibinfo{person}{Eleuterio~F. Toro}.}
  \bibinfo{year}{2009}\natexlab{}.
\newblock \bibinfo{booktitle}{\emph{Riemann {Solvers} and {Numerical} {Methods}
  for {Fluid} {Dynamics}}}.
\newblock \bibinfo{publisher}{Springer Berlin Heidelberg}.
\newblock
\urldef\tempurl%
\url{https://doi.org/10.1007/b79761}
\showDOI{\tempurl}


\bibitem[\protect\citeauthoryear{Ulrich, Gabriel, and Madden}{Ulrich
  et~al\mbox{.}}{2021}]%
        {ulrich20}
\bibfield{author}{\bibinfo{person}{Thomas Ulrich}, \bibinfo{person}{Alice-Agnes
  Gabriel}, {and} \bibinfo{person}{Elizabeth Madden}.}
  \bibinfo{year}{2021}\natexlab{}.
\newblock \showarticletitle{Stress, rigidity and sediment strength control
  megathrust earthquake and tsunami dynamics}.
\newblock \bibinfo{journal}{\emph{Nat. Geosci.}} (\bibinfo{year}{2021}).
\newblock
\urldef\tempurl%
\url{https://doi.org/10.31219/osf.io/9kdhb}
\showDOI{\tempurl}


\bibitem[\protect\citeauthoryear{Ulrich, Vater, Madden, Behrens, van Dinther,
  Van~Zelst, Fielding, Liang, and Gabriel}{Ulrich et~al\mbox{.}}{2019}]%
        {ulrich_coupled_2019}
\bibfield{author}{\bibinfo{person}{Thomas Ulrich}, \bibinfo{person}{Stefan
  Vater}, \bibinfo{person}{Elizabeth~H. Madden}, \bibinfo{person}{J{\"o}rn
  Behrens}, \bibinfo{person}{Ylona van Dinther}, \bibinfo{person}{Iris
  Van~Zelst}, \bibinfo{person}{Eric~J. Fielding}, \bibinfo{person}{Cunren
  Liang}, {and} \bibinfo{person}{A.-A. Gabriel}.}
  \bibinfo{year}{2019}\natexlab{}.
\newblock \showarticletitle{Coupled, physics-based modeling reveals earthquake
  displacements are critical to the 2018 Palu, Sulawesi tsunami}.
\newblock \bibinfo{journal}{\emph{Pure Appl. Geophys.}} \bibinfo{volume}{176},
  \bibinfo{number}{10} (\bibinfo{year}{2019}), \bibinfo{pages}{4069--4109}.
\newblock


\bibitem[\protect\citeauthoryear{Uphoff and Bader}{Uphoff and Bader}{2020}]%
        {uphoff_yet_2019}
\bibfield{author}{\bibinfo{person}{Carsten Uphoff} {and}
  \bibinfo{person}{Michael Bader}.} \bibinfo{year}{2020}\natexlab{}.
\newblock \showarticletitle{Yet Another Tensor Toolbox for Discontinuous
  {Galerkin} Methods and Other Applications}.
\newblock \bibinfo{journal}{\emph{ACM Trans. Math. Softw.}}
  \bibinfo{volume}{46}, \bibinfo{number}{4}, Article \bibinfo{articleno}{34}
  (\bibinfo{date}{Oct.} \bibinfo{year}{2020}), \bibinfo{numpages}{40}~pages.
\newblock
\showISSN{0098-3500}
\urldef\tempurl%
\url{https://doi.org/10.1145/3406835}
\showDOI{\tempurl}


\bibitem[\protect\citeauthoryear{Uphoff, Rettenberger, Bader, Madden, Ulrich,
  Wollherr, and Gabriel}{Uphoff et~al\mbox{.}}{2017}]%
        {uphoff_extreme_2017}
\bibfield{author}{\bibinfo{person}{Carsten Uphoff}, \bibinfo{person}{Sebastian
  Rettenberger}, \bibinfo{person}{Michael Bader}, \bibinfo{person}{Elizabeth~H.
  Madden}, \bibinfo{person}{Thomas Ulrich}, \bibinfo{person}{Stephanie
  Wollherr}, {and} \bibinfo{person}{Alice-Agnes Gabriel}.}
  \bibinfo{year}{2017}\natexlab{}.
\newblock \showarticletitle{Extreme scale multi-physics simulations of the
  tsunamigenic 2004 {Sumatra} megathrust earthquake}. In
  \bibinfo{booktitle}{\emph{Proc. Int. Conf. High Perf. Comput., Networking,
  Storage and Analysis}} \emph{(\bibinfo{series}{{SC} '17})}.
  \bibinfo{publisher}{Association for Computing Machinery},
  \bibinfo{address}{New York, NY, USA}, \bibinfo{pages}{1--16}.
\newblock
\showISBNx{978-1-4503-5114-0}
\urldef\tempurl%
\url{https://doi.org/10.1145/3126908.3126948}
\showDOI{\tempurl}


\bibitem[\protect\citeauthoryear{Verner}{Verner}{2013}]%
        {VernerRungeKutta}
\bibfield{author}{\bibinfo{person}{Jim Verner}.}
  \bibinfo{year}{2013}\natexlab{}.
\newblock \bibinfo{title}{Jim Verner's Refuge for Runge-Kutta Pairs}.
\newblock \bibinfo{howpublished}{\url{http://people.math.sfu.ca/~jverner/}}.
\newblock
\newblock
\shownote{Accessed: 2020-04-01.}


\bibitem[\protect\citeauthoryear{Wilcox, Stadler, Burstedde, and
  Ghattas}{Wilcox et~al\mbox{.}}{2010}]%
        {wilcox_high-order_2010}
\bibfield{author}{\bibinfo{person}{Lucas~C. Wilcox}, \bibinfo{person}{Georg
  Stadler}, \bibinfo{person}{Carsten Burstedde}, {and} \bibinfo{person}{Omar
  Ghattas}.} \bibinfo{year}{2010}\natexlab{}.
\newblock \showarticletitle{A high-order discontinuous {Galerkin} method for
  wave propagation through coupled elastic–acoustic media}.
\newblock \bibinfo{journal}{\emph{J. Comput. Phys.}} \bibinfo{volume}{229},
  \bibinfo{number}{24} (\bibinfo{date}{Dec.} \bibinfo{year}{2010}),
  \bibinfo{pages}{9373--9396}.
\newblock
\showISSN{00219991}
\urldef\tempurl%
\url{https://doi.org/10.1016/j.jcp.2010.09.008}
\showDOI{\tempurl}


\bibitem[\protect\citeauthoryear{Wollherr, Gabriel, and Uphoff}{Wollherr
  et~al\mbox{.}}{2018}]%
        {Wollherr2018}
\bibfield{author}{\bibinfo{person}{Stephanie Wollherr},
  \bibinfo{person}{Alice-Agnes Gabriel}, {and} \bibinfo{person}{Carsten
  Uphoff}.} \bibinfo{year}{2018}\natexlab{}.
\newblock \showarticletitle{{Off-fault plasticity in three-dimensional dynamic
  rupture simulations using a modal Discontinuous {G}alerkin method on
  unstructured meshes: implementation, verification and application}}.
\newblock \bibinfo{journal}{\emph{Geophys. J. Int.}} \bibinfo{volume}{214},
  \bibinfo{number}{3} (\bibinfo{date}{sep} \bibinfo{year}{2018}),
  \bibinfo{pages}{1556--1584}.
\newblock
\showISSN{0956-540X}
\urldef\tempurl%
\url{https://doi.org/10.1093/gji/ggy213}
\showDOI{\tempurl}


\bibitem[\protect\citeauthoryear{Wylie}{Wylie}{2020}]%
        {Wylie2020}
\bibfield{author}{\bibinfo{person}{Brian J.~N. Wylie}.}
  \bibinfo{year}{2020}\natexlab{}.
\newblock \showarticletitle{Exascale potholes for {HPC}: Execution performance
  and variability analysis of the flagship application code {HemeLB}}. In
  \bibinfo{booktitle}{\emph{2020 {IEEE}/{ACM} Int.\ Workshop {HPC} User Support
  Tools ({HUST}) \& Workshop Program.\ \& Perform.\ Visual.\ Tools
  ({ProTools})}}. \bibinfo{publisher}{{IEEE}}.
\newblock
\urldef\tempurl%
\url{https://doi.org/10.1109/hustprotools51951.2020.00014}
\showDOI{\tempurl}


\bibitem[\protect\citeauthoryear{Zhan, Cantono, Kamalov, Mecozzi, M{\"u}ller,
  Yin, and Castellanos}{Zhan et~al\mbox{.}}{2021}]%
        {Zhan2021}
\bibfield{author}{\bibinfo{person}{Zhongwen Zhan}, \bibinfo{person}{Mattia
  Cantono}, \bibinfo{person}{Valey Kamalov}, \bibinfo{person}{Antonio Mecozzi},
  \bibinfo{person}{Rafael M{\"u}ller}, \bibinfo{person}{Shuang Yin}, {and}
  \bibinfo{person}{Jorge~C. Castellanos}.} \bibinfo{year}{2021}\natexlab{}.
\newblock \showarticletitle{Optical polarization{\textendash}based seismic and
  water wave sensing on transoceanic cables}.
\newblock \bibinfo{journal}{\emph{Science}} \bibinfo{volume}{371},
  \bibinfo{number}{6532} (\bibinfo{year}{2021}), \bibinfo{pages}{931--936}.
\newblock
\urldef\tempurl%
\url{https://doi.org/10.1126/science.abe6648}
\showDOI{\tempurl}


\end{thebibliography}

\end{document}